Banner appropriate to article type will appear here in typeset article

# Bubble nucleation and jetting inside a millimetric droplet

**Juan Manuel Rosselló**[1,2]†, **Hendrik Reese**[1], **K. Ashoke Raman**[1], and **Claus-Dieter Ohl**[1]

[1]Otto von Guericke University Magdeburg, Institute of Physics, Universitätsplatz 2, 39106 Magdeburg, Germany.

[2]Faculty of Mechanical Engineering, University of Ljubljana, Aškerčeva 6, 1000 Ljubljana, Slovenia.

In this work, we present experiments and simulations on the nucleation and successive dynamics of laser-induced bubbles inside liquid droplets in free-fall motion, i.e. a case where the bubbles are subjected to the influence of a free boundary in all directions. The droplets of a millimetric size are released from a height of around 20 cm and acquire a nearly spherical shape by the moment the bubble is nucleated. Within this droplet, we have investigated the nucleation of secondary bubbles induced by the rarefaction wave that is produced when the shock wave emitted by the laser-induced plasma reflects at the drop surface. Interestingly, three-dimensional clusters of cavitation bubbles are observed. Their shape is compared with the negative pressure distribution computed with a CFD model and allows us to estimate a cavitation threshold value. In particular, we observed that the focusing of the waves in the vicinity of the free surface can give rise to explosive cavitation events that end up in fast liquid ejections. High-speed recordings of the drop/bubble dynamics are complemented by the velocity and pressure fields simulated for the same initial conditions. The effect of the proximity of a curved free surface on the jetting dynamics of the bubbles was qualitatively assessed by classifying the cavitation events using a non-dimensional stand-off parameter $\Upsilon$ which depends on the drop size, the bubble maximum radius and the relative position of the bubble inside the drop. Additionally, we studied the role of the drop's curvature by implementing a structural similarity algorithm to compare cases with bubbles produced near a flat surface to the bubbles inside the drop. Interestingly, this quantitative comparison method indicated the existence of equivalent stand-off distances at which bubbles influenced by different boundaries behave in a very similar way. The oscillation of the laser-induced bubbles promote the onset of Rayleigh-Taylor and Rayleigh-Plateau instabilities, observed on the drop's surface. This phenomenon was studied by varying the ratio of the maximum radii of the bubble and the drop. The specific mechanisms leading to the destabilisation of the droplet surface were identified through a careful inspection of the high speed images together with the numerical simulations.

† Email address for correspondence: jrossello.research@gmail.com

Abstract must not spill onto p.2

## 1. Introduction

Phase explosion in confined liquid volumes has recently gained interest because of its connection with thriving research areas like x-ray liquid crystallography (Grünbein *et al.* 2021), x-ray holography (Vassholz *et al.* 2021; Hagemann *et al.* 2021), extreme UV light, and plasma generation (Favre *et al.* 2002). A better understanding of the interaction of high-power lasers with small liquid particles is also relevant in laser-based atmospheric monitoring techniques (Rohwetter *et al.* 2010; Mei & Brydegaard 2015) or in optical atomisation techniques that can be applied to the production of airborne transported micro-drops used as drug carriers (Lee *et al.* 2022). At the heart of all of these research fields is the injection of high-power photons into a small liquid sample, the initiation of phase transition from liquid to vapour, the rapid pressure fluctuations, and the successive complex fluid mechanics driven by this impulsive energy input. In this study, we want to shed light on the fundamental flows that can be induced in liquid samples once this phase transition has been initiated. In particular, we focus on the fluid dynamics within a spherically confined liquid sample after the violent phase explosion of the vapour bubble induced by a high-power laser pulse. We explore the non-spherical dynamics of vapour bubbles within a liquid droplet, i.e. surrounded by free boundaries only. Bubble dynamics in droplets have so far mostly been studied from the perspective of destabilisation of the liquid-gas interface of the droplet (Singh & Knight 1980; Alexander & Armstrong 1987; Eickmans *et al.* 1987; Lindinger *et al.* 2004; Thoroddsen *et al.* 2009; Marston & Thoroddsen 2015; Gonzalez-Avila & Ohl 2016; Zeng *et al.* 2018). Here, we explore the bubble dynamics within the droplet (Obreschkow *et al.* 2006).

Pulsed lasers can be focused into optically transparent media to induce explosive bubble nucleation by dielectric breakdown. This process is accompanied by the emission of an acoustic shock wave with an amplitude on the order of GigaPascals depending on the pulse energy, duration, and wavelength. For instance, the initial amplitude of the shock wave (i.e. at the edge of the plasma rim) in water can be in the range from 2.4 GPa to 11.8 GPa for a 6 ns laser pulse with an energy between 1 mJ to 10 mJ and a wavelength of 1064 nm focused with a numerical aperture (NA) of 22° (Vogel *et al.* 1996; Noack & Vogel 1998). Recently, the initial shock wave amplitude produced by similar nanosecond laser pulses of 24 mJ (NA = 10°) was measured with a novel x-ray probing technique, obtaining peak values of around 20 GPa (Vassholz *et al.* 2021).

When a laser-induced cavity is produced in a confined space with free boundaries, like a droplet, most of the sound wave energy reflects back from the interface with an inverted phase, meaning that the original shock wave is transformed into a rarefaction wave. If the negative pressure amplitude of the reflected wave is below the cavitation threshold of the liquid, a trail of bubbles is nucleated after the wave passage. This effect is commonly observed upon wave reflection on the free boundary of a flat surface (Heijnen *et al.* 2009), nearby bubbles (Quinto-Su & Ando 2013), a liquid column (Sembian *et al.* 2016) or, as we already mentioned, a drop (Obreschkow *et al.* 2006; Gonzalez-Avila & Ohl 2016; Kondo & Ando 2016; Kyriazis *et al.* 2018; Wu *et al.* 2018*b*, 2021; Biasiori-Poulanges & Schmidmayer 2023). Laser cavitation in some of these configurations was lately applied in studies involving x-ray holography or x-ray diffraction to investigate the propagation of shock waves in liquids (Stan *et al.* 2016; Ursescu *et al.* 2020; Hagemann *et al.* 2021). The use of very small amounts of liquid prevents the x-rays from being fully absorbed by the sample, thus improving the contrast of the x-ray images. This technique is suitable to study the properties of opaque liquids without optical aberrations, it is less sensitive to distortions produced by wavy surfaces, and also allows retrieving information about the liquid density changes produced by the passage of the pressure waves (Vassholz *et al.* 2021, 2023), which represents an advantage over traditional optical imaging.

Another interesting aspect of the nucleation of bubbles in the proximity of a boundary resides in their jetting dynamics. Laser-induced bubbles produced under different boundary conditions have been widely studied, both experimentally and numerically. Perhaps the case that got the most attention is the one of a bubble collapsing in the proximity of a boundary of large extent, e.g. a solid boundary (Plesset & Chapman 1971; Lauterborn & Bolle 1975; Blake *et al.* 1999; Brujan *et al.* 2002; Lindau & Lauterborn 2003; Yang *et al.* 2013; Lechner *et al.* 2017; Gonzalez-Avila *et al.* 2021), an elastic boundary (Brujan *et al.* 2001; Rosselló & Ohl 2022), or a free surface (Koukouvinis *et al.* 2016; Li *et al.* 2019*c*; Bempedelis *et al.* 2021; Rosselló *et al.* 2022). In real-world conditions, the boundary is of finite extent and the cavity may be spuriously affected by more than a single boundary (for instance, the walls of a container or the liquid free surface), exerting a considerable influence the direction of the jetting (Kiyama *et al.* 2021; Andrews & Peters 2022).

The jet dynamics are frequently characterised by a stand-off parameter (Lindau & Lauterborn 2003; Supponen *et al.* 2016; Lauterborn *et al.* 2018) computed as the ratio of the distance between the bubble nucleation position and the boundary ($d$) and the maximum radius attained by the bubble after its creation ($R_{max}$). If the cavity collapse occurs next to boundaries other than a plane, for instance, irregular or curved surfaces (Tomita *et al.* 2002; Blake *et al.* 2015; Wu *et al.* 2018*a*; Li *et al.* 2019*b*; Aganin *et al.* 2022) like pillars (Koch *et al.* 2021*b*; Kadivar *et al.* 2021), fibers (Mur *et al.* 2023), corners (Zhang *et al.* 2020; Mahmud *et al.* 2020), crevices (Trummler *et al.* 2020; Andrews *et al.* 2020), perforated plates (Gonzalez-Avila *et al.* 2015; Reese *et al.* 2022), or spheres (Zhang *et al.* 2018; Li *et al.* 2019*a*; Zevnik & Dular 2020; Ren *et al.* 2022), the anisotropy does not have one predominant direction and thus the use of a single stand-off parameter (e.g. $d/R_{max}$) is no longer sufficient to fully characterise the system. The same situation arises in cases where the bubbles are produced in a constricted space, for example in narrow channels (Gonzalez-Avila *et al.* 2011; Wang *et al.* 2018; Brujan *et al.* 2022), between two surfaces (Li *et al.* 2017; Liu *et al.* 2017), in a liquid column (Robert *et al.* 2007), or inside a drop (Obreschkow *et al.* 2006; Thoroddsen *et al.* 2009; Marston & Thoroddsen 2015; Gonzalez-Avila & Ohl 2016; Zeng *et al.* 2018).

The dynamics of jetting bubbles inside drops or curved free surfaces have not been extensively explored. Recently, we have reported experimental and numerical results on the formation of a jetting bubble in the proximity of a curved free boundary, given by the hemispherical top of a water column or a drop sitting on a solid plate (Rosselló *et al.* 2022). As a natural extension of that work, we now present a study on the jet formation during the collapse of laser-induced bubbles inside a falling drop. This is a particularly interesting case as the bubble is surrounded entirely by a free boundary. From an experimental point, the intrinsic curvature of the liquid surface offers a very clear view into the bubble's interior.

The rapid acceleration induced by the bubble oscillations in the proximity of a free boundary also gives rise to surface instabilities, in particular Rayleigh-Taylor instabilities (RTI) (Taylor 1950; Keller & Kolodner 1954; Zhou 2017*a,b*). This situation is more pronounced when the oscillating bubble wall gets close to the free surface, as commonly occurs in reduced volumes like a drop (Zeng *et al.* 2018; Klein *et al.* 2020). The Rayleigh-Taylor instability produces corrugated patterns on the liquid surface that can grow and promote the onset of other instabilities like the Rayleigh-Plateau instability. Furthermore, the multiple pits and ripples produced by the RTI on the liquid surface can interact with the acoustic emissions of the oscillating bubble to generate a fluid focusing which results in a thin outgoing liquid jet (Tagawa *et al.* 2012; Peters *et al.* 2013).

This article is organised into different sections focusing on one of the above-discussed aspects, i.e. the shock wave dynamics and the nucleation of secondary cavitation bubbles, the jetting dynamics of the collapsing laser-induced bubbles, and the formation of instabilities on the drop surface as a consequence of the bubble oscillation.

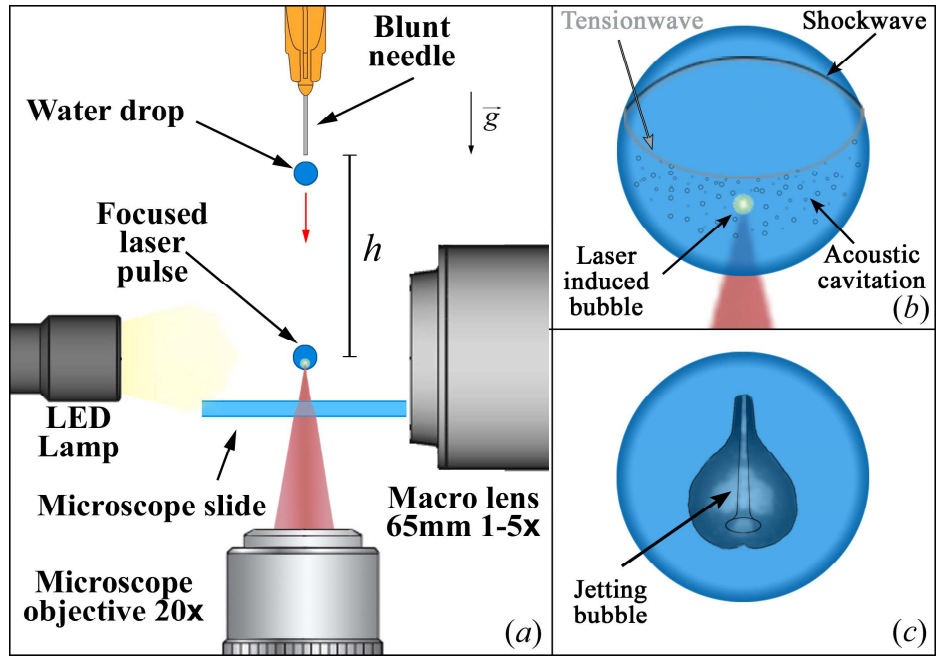

Figure 1: Description of the experimental setup. (a) A water drop with a volume of ~ 12 µl is detached from a cylindrical blunt needle (stainless steel, 600 µm of external diameter) by gravitational forces. When the drop reaches a velocity of $(1.7 \pm 0.1)$ m/s a cavitation bubble is produced inside it by a laser pulse with a duration of 4 ns and a typical energy of $(2.4 \pm 0.1)$ mJ. (b) Once reflected from the drop surface, the shock waves emitted from the laser-induced bubble nucleate tiny bubbles inside the liquid drop. (c) The bubble undergoes an asymmetric collapse with jetting, whose shape depends on the position of the bubble inside the drop.

## 2. Experimental method

The experimental method used to achieve controlled laser bubble inception inside a milli-metric drop is depicted in figure 1(a). Individual drops were released from the tip of a blunt metallic needle with an internal diameter of 330 µm (and an external diameter of 600 µm) by the action of an electronic syringe pump *KD Scientific - Legato SPLG110*. This device pushed a fixed volume of ~ 12 µl of deionised water through the needle, producing single drops with a radius of $(1.42 \pm 0.01)$ mm. After a drop was released it traveled a distance of $h = 30$ cm in free-fall motion. Just before it impacted a glass plate a pulsed laser was focused into the droplet to nucleate the cavitation bubble.

The pulse energy of the laser (Nd:YAG *Q2-1064* series, pulse duration 4 ns, wavelength 1064 nm) could be varied between 1.9 mJ and 20.3 mJ and was focused with a microscope objective (*Zeiss LD Achroplan* 20×, NA = 0.4) see bottom of figure 1(a). In the experiments, a standard microscope slide was placed on top of the laser focusing objective in order to prevent wetting of its outer lens, which would provoke a significant distortion of the laser beam. Accordingly, the protective glass was meticulously cleaned after each drop impact.

The fall distance $h$ was sufficient for the surface tension to stabilise the liquid into an approximately spherical shape, reaching a velocity of $(1.7 \pm 0.1)$ m/s upon laser arrival. At the same time, the variation of the lateral position of the drop centre relative to the laser focus was typically below 200 µm, which aids experimental repeatability. The vertical position where the bubble is created within the droplet is controlled with some precision by synchronising the laser pulse with the passage of the drop through a light gate. This consists of a red laser diode paired with a photo-diode that triggers a digital delay generator *Quantum 9520* which then fires the laser after a specified time.

The dynamics of the cavitation bubble within the droplet and the resulting surface instabilities were captured in high-speed videos using a *Shimadzu XPV-X2* camera equipped with a photography macro lens *Canon MP-E 65 mm f/2.8 1-5×*. A diffused back illumination from a continuous white LED lamp *SMETec* (9000 lm) in combination with the curved nature of the drops allowed us to obtain clear images of the droplet interior. Furthermore, the curvature of the liquid refracted the light in a way that reveals the internal structures of the jetting bubbles, knowing that it distorts the apparent position and shape (Koch *et al.* 2021*a*). For direct comparison of the experimental and the numerical results, an *in-house* script was applied to the simulated results to compensate for such image distortions (Martins *et al.* 2018). This correction (based on Snell's law) was also used to obtain the "real" nucleation position of the laser bubble.

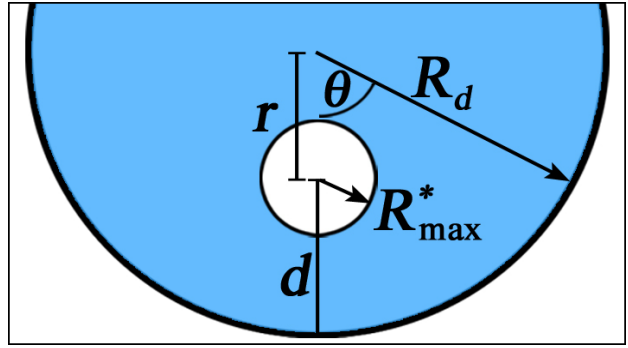

Figure 2: Schematic of the drop interior with relevant dimensional parameters.

Due to the limited number of recorded frames, the framing rate of the high-speed videos had to be adjusted to capture the important features of the phenomena under study. For instance, to visualise the shock wave propagation and the resulting nucleation of bubbles from the reflected rarefaction wave, see figure 1(b)) required a frame rate of 5 Mfps (i.e. the maximum achievable by the camera), while the temporal evolution of the jets (depicted in figure 1(c)) and the instabilities of the drop surface are captured already at 200 kfps or 500 kfps, respectively.

### 2.1. *Definition of a stand-off parameter for a curved boundary* ϒ

In order to consider the curvature of the drop's surface in the characterisation of the jet dynamics, we defined a non-dimensional coefficient ϒ that combines two non-dimensional numbers, each one representing a relevant dimension of the problem. First, we use the stand-off distance $D^*$ (Lauterborn *et al.* 2018) as the ratio of the bubble "seeding" position ($d$) and the maximum radius achieved by the bubble when produced at the centre of the spherical drop ($R^*_{max}$). The second non-dimensional distance $\chi$ is given by the ratio of the drop radius ($R_d$) and the distance of the bubble from the drop centre ($r$). To summarise,

$$D^* = \frac{d}{R^*_{max}} \tag{2.1}$$

$$\chi = \frac{R_d}{r} = \frac{R_d}{R_d - d} \tag{2.2}$$

$$\Upsilon = \chi \, D^* = \frac{R_d}{(R_d - d)} \, \frac{d}{R^*_{max}} \tag{2.3}$$

A schematic representation of the aforementioned parameters is presented in figure 2. Here, $R^*_{max}$ is tightly related to the energy of the laser pulse (Lauterborn *et al.* 2018) and, as we explain later in section 4.3, it also varies slightly with the drop size as $R_d \to \infty$. For the purpose of having reproducible results, the use of ϒ should be limited to values of $R^*_{max}$ for which the bubble is contained inside the drop volume (i.e. $0 \leq R^*_{max} < R_d$) and the drop shape is not significantly distorted by surface instabilities (Zeng *et al.* 2018). Additionally, the symmetry of the drop/bubble configuration implies that $d \leq R_d$.

In principle, the parameter $\Upsilon$ behaves similarly as the traditional stand-off distance (e.g. $d/R_{max}$), however, the addition of $\chi$ as a weighting factor represents a measure of the influence of the boundaries all around the bubble, and not only its closest point. This means that the regions of the free surface in directions other than $\theta = 0$ could also be relevant to the bubble dynamics as the separation from the bubble and the boundary in those angular directions gets smaller, i.e. when the radius $R_d$ is decreased and the bubble is located at a reduced $d$. Alternatively, $\Upsilon$ could be understood as a measure of the anisotropy, with high anisotropy at the liquid boundary ($d \to 0$) and perfect isotropy at the bubble centre, $d = R_d$.

The tight relation between the traditional stand-off distance and $\Upsilon$ is also evidenced by the following considerations and limiting cases:

- $\Upsilon$ rises monotonically with $d$ for a fixed laser pulse energy (or $R^*_{max}$).
- In the limit $R_d \to \infty$ the traditional stand-off distance is recovered.
- If the bubble is near the drop wall, $d \to 0$, then $\Upsilon \to 0$.
- If the bubble is near the drop centre, $d \to R_d$, then $r = 0$, $\Upsilon \to \infty$, and we recover the traditional unbounded case, in which the bubble collapses spherically due to symmetry.

It is important to note that $\Upsilon$ can take the same value for different combinations of $d$, $R^*_{max}$, and $R_d$. Therefore, two identical values of $\Upsilon$ computed from two different values of $D^*$ and $\chi$ do not necessarily result in identical bubble dynamics. A comparison between cases could be made by fixing the value of one or two of the parameters. For example, the effect of the surface curvature $R_d$ on the bubble dynamics can be evaluated by maintaining $D^*$, or the influence of the "seeding" depth $d$ can be studied by fixing the drop size $R_d$ and the energy of the laser pulse. In this way, the parameter preserves the same functionality as the traditional stand-off parameter (Lauterborn *et al.* 2018), but now includes the surface curvature dimension.

## 3. Numerical method

Volume-of-Fluid simulations were carried out in *OpenFOAM-v2006* (OpenFOAM-v2006 2020) using a modification of the solver *compressibleMultiphaseInterFoam*. This modified version is called *MultiphaseCavBubbleFoam* and was already implemented in previous works to study the formation of the "bullet jet" (Rosselló *et al.* 2022) and micro-emulsification (Raman *et al.* 2022). In those works, similar simulations of a single expanding and collapsing bubble in the vicinity of a liquid-gas and a liquid-liquid interface were performed, respectively. Since the solver is explained in detail there, we will only give the information that is specific to the present case of a bubble created in a free-falling liquid drop.

Considering the approximate rotational symmetry of the experimental configuration, we carried out the simulations as quasi-two-dimensional. The computational domain represents a slice of a cylindrical domain with a height of 3 mm and a radius of 3 mm, which is filled with a gas representing the surrounding air at ambient pressure. The domain is divided into a square mesh of cells with a width of 40 µm, which is then further refined to a cell width of 10 µm in the region occupied by the liquid drop. The boundaries of the domain in the radial and axial directions are open, wave transmissive boundaries.

A slightly prolate ellipsoidal liquid drop representing a falling water drop is initiated in the centre of the cylinder with an axial radius of 1440 µm and a radial radius of 1400 µm. We neglect the relative motion of the drop through the air, and thus take the drop and the air to be initially at rest. This is because the speed of the falling drop and the effects of drag are negligible when compared with the speeds developed by the bubble wall and the jets. We also neglect any subsequent gravitational acceleration, since its effect is negligible on the time scales considered. Inside the drop, a bubble is seeded on the symmetry axis with an initial over-pressure of 1.69 GPa and an initial radius of 25.7 μm. The initial pressure was chosen such that the initial bubble gas density equals the density of the surrounding liquid, in accordance with equation (3.1). This is based on the assumption that the laser energy deposition occurs on a much smaller time scale than the expansion of the bubble. The initial bubble radius $R_0$ is chosen to match the maximum expansion $R^*_{max}$ in the experiment.

|          | $B$ in MPa | $\rho_0$ in kg/m$^3$ | $p_0$ in Pa | $\gamma$ | $\mu$ in mPa s |
|----------|------------|----------------------|-------------|----------|----------------|
| liquid   | 303.6      | 998.2061             | 101325      | 7.15     | 1              |
| gases    | 0          | 0.12                 | 10320       | 1.33     | 0.013          |

Table 1: Tait equation of state parameters and dynamic viscosities $\mu$ of the simulated fluid components. Both gaseous components are treated as the same type of gas.

The bubble contents are modelled with the same properties as the gas surrounding the liquid droplet but are calculated as a separate component. This allows us to apply a mass correction to the gas in the bubble only that accounts for the mass loss due to condensation during the bubble's first oscillation cycle. This is done as a one-time correction at the time of maximum bubble expansion, at which the bubble gas density is reduced by 70 %. More details can be found in our previous work (Rosselló *et al.* 2022). The surface tension between the liquid and the gases is 70 mN/m, and that between the gases is 0. The Tait equation of state is used for all components,

$$p = (p_0 + B) \left( \frac{\rho}{\rho_0} \right)^{\gamma} - B \,, \qquad (3.1)$$

with the parameters given in table 1. Here, $\gamma$ is the adiabatic exponent.

The output of the numerical data was done in intervals of 10 ns to capture shock wave propagation dynamics, and every 1 μs for the bubble and jetting dynamics.

## 4. Results and discussion

The inception of a laser-induced bubble inside a liquid drop gives rise to a rich and complex chain of events. We start with an overview of the fluid dynamics that are observed following the creation of the cavitation bubble by the dielectric rupture of the liquid, as shown in figure 3. Here, the bubble is nucleated off-centre and close to the upper interface of the droplet. The fluid dynamics can be divided into three stages, which are discussed in detail in the later sections. For now, we provide a brief description of these 3 stages: (1) The bubble is nucleated into a rapidly expanding vapour cavity that launches during its deceleration a shock wave into the droplet, not visible in figure 3. Upon reflection at the acoustic soft liquid-gas interface, the rarefaction wave propagates through the drop leaving behind a trail of cavitation bubbles in certain regions where the wave convergence produces sufficient tension to induce local acoustic cavitation, $2 \, \mu s \leq t \leq 6 \, \mu s$ in figure 3. Depending on the location of the laser bubble the rarefaction wave may focus in a reduced volume close to the interface, creating secondary cavitation and provoking the ejection of a single jet at the opposite side of the laser bubble nucleation site (e.g. $t > 6 \, \mu s$ in figure 3). (2) In the second stage, the laser-induced bubble undergoes an asymmetrical collapse from its maximum size. Here, the anisotropy of the boundary conditions results in the formation of a jet, which starts as an indentation on one side of the cavity and grows to pierce the bubble at the opposite extreme. In cases where the laser cavity is created near the drop surface, we also observe the destabilisation of the liquid surface by a Rayleigh-Taylor instability. (3) In the third and last stage, the bubble re-expands after jetting, adopting a liquid-gas structure that depends mostly on the stand-off distance (i.e. Υ). On its second collapse, the cavity fragments and later disperses due to the complex flow created by its first collapse.

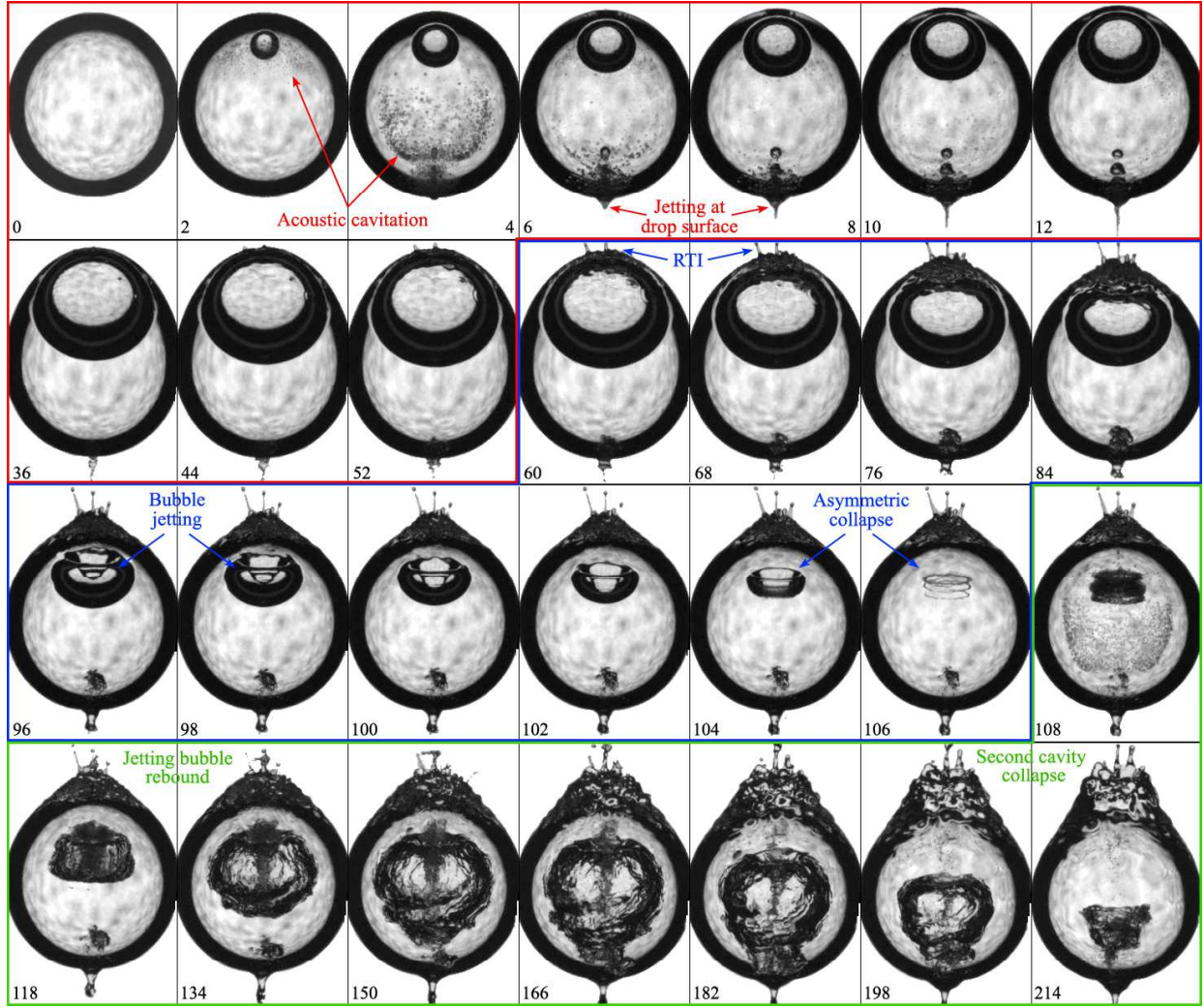

Figure 3: Stages of the events developing inside the drop. The numbers indicate the time in µs after the laser shot. In the first stage, spanning from $t = 0$ µs to $t = 52$ µs(framed in red), a rarefaction wave (i.e. the reflection of the shock wave) produces a trail of cavitation bubbles. For low values of Υ a liquid jet is ejected from the extreme of the drop opposite to the bubble inception. In the second stage, defined between $t = 60$ µs and $t = 106$ µs(framed in blue), the bubble collapses after reaching its maximum size and a jet forms. In some cases, a Rayleigh-Taylor instability (RTI) is observed near the bubble. The third stage (framed in green) runs from $t = 108$ µs until the end of the video at $t = 214$ µs. Here, the bubble re-expands after jetting and adopts a characteristic shape during its second collapse that depends mostly on Υ. The width of each frame is 2.70 mm.

In the following, the reported values of Υ are computed for a surface curvature of 1.42 mm, which corresponds to the mean radius of the drops produced in this work.

### 4.1. *Acoustic cavitation nucleation*

The specific shape of the cavitation bubble clusters produced by the passage of the rarefaction wave is highly dependent on Υ. This is because the negative pressure focuses differently when the original shock wave is emitted from a different location. As the acoustic nucleation only occurs below a certain pressure threshold, the resulting bubble clouds can assume complex three-dimensional structures. Figure 4 presents experimental results showing the temporal evolution of bubble clouds generated for different values of $\Upsilon$. In this study, the bubble "seeding" position was varied by changing the delay between the drop release and the laser shot, thus shifting the laser focus position along the vertical symmetry axis of the drop.

As aforementioned, the shock waves emitted from the laser focal spot will reflect from the free boundary of the drop as a rarefaction wave. Due to the nearly spherical shape of the drop, the reflected acoustic waves will focus in a region located at a similar distance from its centre $r$ (where the laser bubble was created) but on the opposite side of the drop. In the case where the shock wave originates near the surface (i.e. $\Upsilon \lesssim 1$), the resulting pressure distribution is characterized by a negative pressure zone moving close to the liquid surface which produces a spherical shell of tiny cavitation bubbles, as displayed in the panels (a) to (c) (and also (i) to (j)) of figure 4. This phenomenon occurs when the sound reflects multiple times on the drop walls and travels circumferentially near the liquid surface without a significant loss of intensity, which is usually referred to as "whispering gallery effect" (Raman & Sutherland 1922). As the rarefaction waves focus at a similar depth where the shock wave was emitted, it produces explosive cavitation events close to the free boundary and on the drop's vertical axis. The rapid expansion of those larger cavitation bubbles gives rise to the liquid jets shown in the first row of figure 3. A more detailed explanation of the formation and dynamics of this particular type of jet will be published elsewhere.

As the laser focusing depth $d$ is increased, the negative pressure is distributed in larger regions, but still, the nucleation of bubbles predominantly occurs on the side opposite to the laser focus. Additionally, the bubble clusters turn from having the structure of a shell (see panels (g), (h), and (i) of figure 4) into a volumetric cavitation cloud when the laser bubble is generated near the drop centre, as shown in the panels (e) and (f) of figure 4. This transition can be explained by analysing the pressure distribution dynamics with the numerical simulations (Ando *et al.* 2012; Quinto-Su & Ando 2013; Gonzalez-Avila & Ohl 2016). Figure 5 demonstrates the clear correlation between the evolution of the acoustic pressure profile and the nucleation of secondary cavitation bubbles. Furthermore, this correlation can be used to determine the cavitation pressure threshold of the liquid by comparing the shape and the location of the negative pressure front with the shape of the bubble cloud within the drop. Such a comparison was only possible after applying a numerical algorithm to the simulated results to compensate for the image distortions induced by the drop curvature. The last frames in panels (a) and (b) of figure 5 display an overlap of both the experimental video frames and the simulated pressure profiles. From the measurements, we found a consistent cavitation threshold of approximately 4.5 MPa. Considering that we did not filter the water sample we assume that the cavitation is most likely heterogeneous.

The acoustic cavitation thresholds reported for water in the literature vary strongly, depending on the measurement method, water purity, gas saturation, and water temperature. Atchley *et al.* (1988) used distilled, deionised, and filtered (0.2 μm) tap water irradiated by pulsed ultrasound and found thresholds between 0.5 and 2.0 MPa, depending on the pulse duration and frequency. Sembian *et al.* (2016) subjected a water column to a single shock wave and found a cavitation threshold between 0.42 and 2.33 MPa. Biasiori-Poulanges & Schmidmayer (2023) compared numerical simulations and experiments of a liquid drop subjected to a planar shock wave and found a threshold between 0.37 and 2.4 MPa. A similar shock front can be found when a droplet impacts on a solid surface at a high speed (e.g. higher than 100 m/s) as studied by Kondo & Ando (2016); Wu *et al.* (2018*b*, 2021). Assuming homogeneous nucleation, Ando *et al.* (2012) and later Quinto-Su & Ando (2013) found a cavitation threshold of 60 MPa and 20 MPa, respectively, comparing experiments and simulations of a reflected shock wave at a free boundary. Therefore, the threshold value obtained in this work falls around the middle of the spectrum of values measured by other authors. Figure 5(c) evidences a growth in the secondary bubble cluster with increasing energy of the laser pulse, demonstrating the resulting shift in the location of the cavitation threshold isobar for higher amplitudes of the initial shock wave. It is relevant to point out that VoF simulations are notorious for numerical diffusion which causes the shock wave to smear out over time. Because of this, the simulations may underestimate the pressures reached in the experiments. Please note that the VoF model does not account for phase transitions and the subsequent interaction of nucleated cavitation bubbles with the finite amplitude waves. A model for of high-frequency waves interacting with small cavitation clouds that may be applicable was recently developed by Maeda & Colonius (2019). Finally, panels (d) and (e) of figure 5 exemplify some of the hollow three-dimensional bubble structures observed in the experiments.

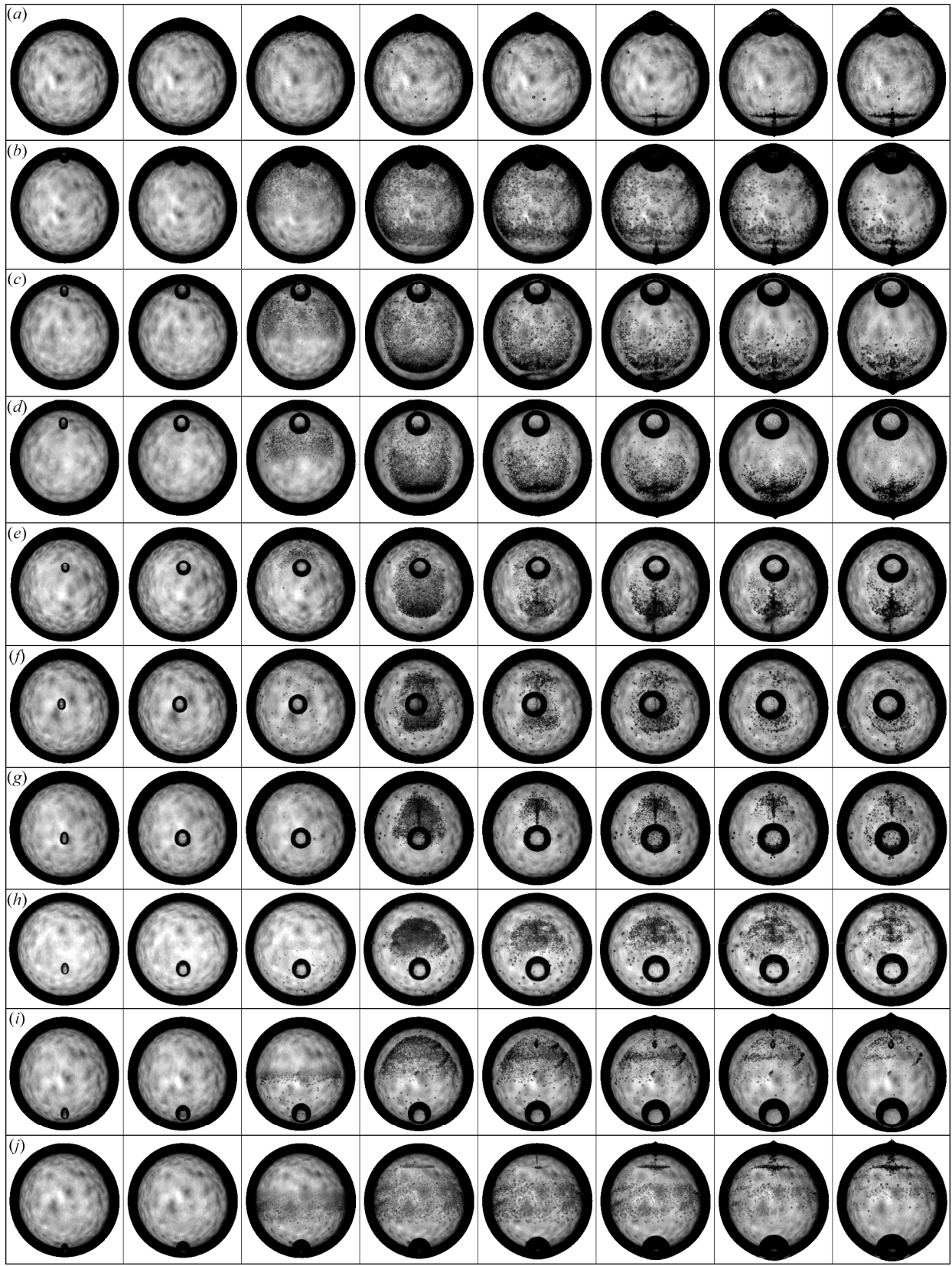

Figure 4: Acoustic cavitation inside a water droplet. The distribution of bubbles in the liquid changes significantly with the position of the laser-induced bubble. The frame width is 3.15 mm. The time between consecutive frames is 600 ns. (a) $\yen = 0.65$. (b) $\yen = 1.1$. (c) $\yen = 1.7$. (d) $\yen = 2.5$. (e) $\yen = 7.5$. (f) $\yen = 68$. (g) $\yen = 13$. (h) $\yen = 5.4$. (i) $\yen = 1.8$. (j) $\yen = 0.9$. Full videos of panels (b), (d) and (f) are available in the online supplementary movies 1-3.

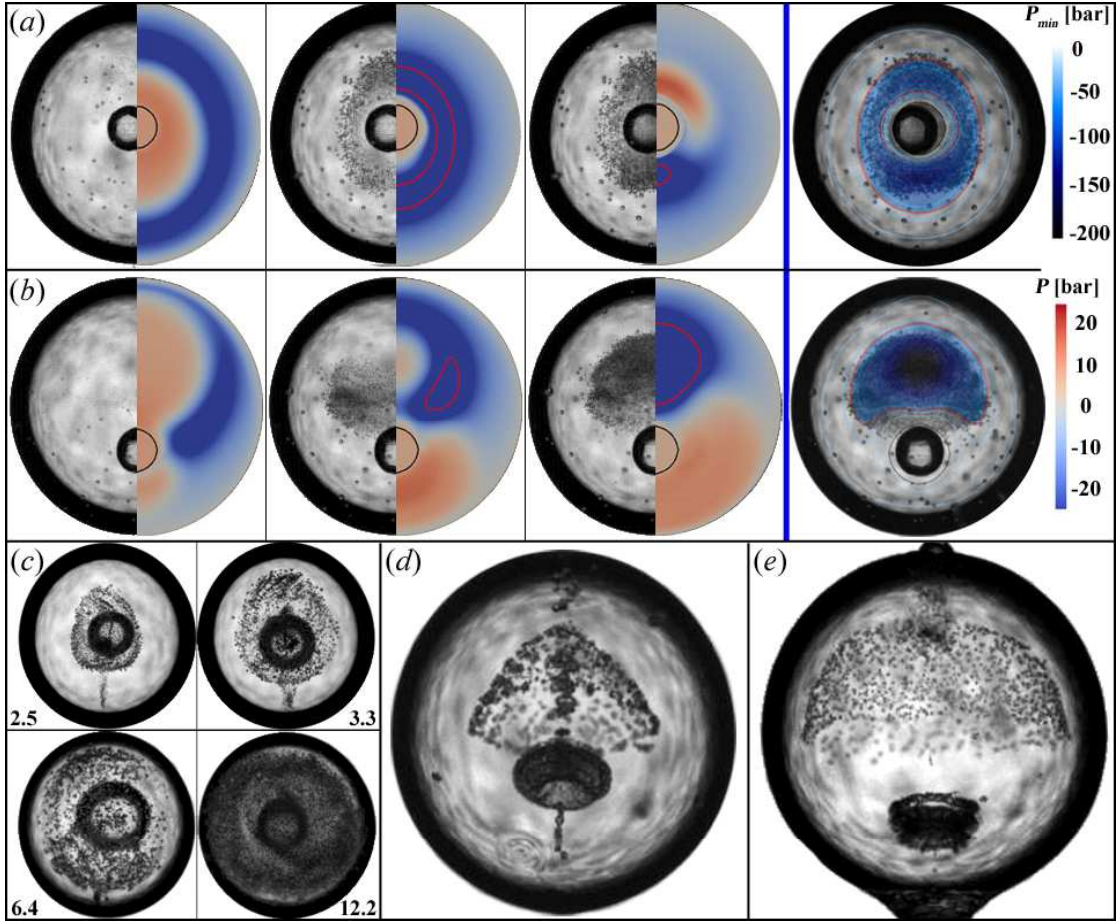

Figure 5: Acoustic cavitation bubble clouds for laser-induced bubbles at different relative positions in the drop. The frames compare the advance of the shock/tension waves within the drop with the observed nucleation sites. The average drop diameter is $(2.84 \pm 0.05)$ mm in all cases. The last frame of each series presents an overlay of the frames and the cumulative minimum pressure after the first reflection of the shock wave at the free boundary. The red line indicates the isobar of -4.5 MPa, i.e. the approximate nucleation threshold pressure. (a) Here, the bubble is slightly off-centre (i.e. $d \simeq R_d$). (b) ¥ = 5.4. (c) Change in the cluster dimensions with increasing laser pulse energy (indicated in mJ). (d) and (e) present evidence of the formation of complex hollow three-dimensional bubble structures. Here, ¥ is 3.5 and 1.45, respectively.

### 4.2. *Bubble jetting*

In the second stage presented in figure 3 the laser-induced bubble reaches its maximum radius and then collapses. At this point, it becomes clear that a non-uniform distance between the bubble and the free surface produces an asymmetric collapse, which culminates in a liquid jet. In this section, we explore the effect of varying the parameter Υ (as performed in section 4.1), but this time we lay focus on the development of the jets, as shown in figure 6.

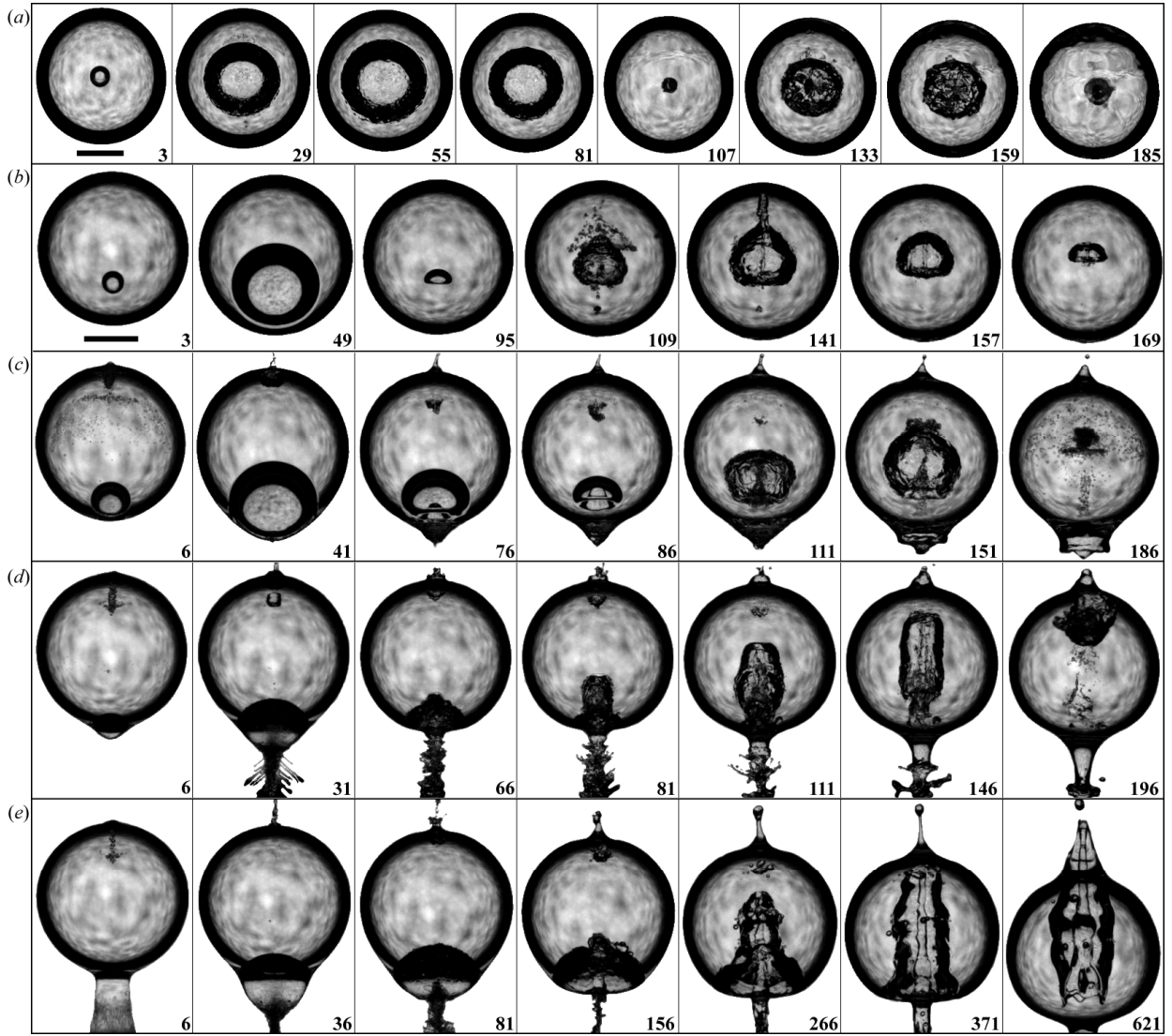

Figure 6: Bubble jetting is produced by a laser-induced bubble generated at different relative positions inside the drop. The numbers indicate the time in μs. The length of the scale bars is 1 mm. (a) Spherical oscillation case, Υ = 203. (b) Weak jet case, Υ = 3.9. (c) Standard jet case, Υ = 1.5. (d) Υ = 0.44. (e) Bullet jet case, Υ = 0.22. Full videos are available in the online supplementary movies 4-8.

The experiments reveal that, as the position of the laser focus is varied between the centre and the surface of the drop, the characteristics of the jetting change smoothly: For large values of Υ, a spherical rebound of the bubble without any jetting is observed. The values of Υ ≳ 3.5 are accompanied by the formation of a very thin liquid jet crossing through the centre of a weakly deformed bubble. In this "weak jet" case, the tip of the jet separates from the main cavity when it starts to collapse during its second oscillation cycle (see figure 6(b)). For 1.2 ≲ Υ ≲ 3.5, as in panel (c) of figure 6, the "whispering gallery" effect becomes relevant, causing the inception of larger acoustic bubbles on the side opposite to the laser cavity and the ejection of liquid driven by their expansion. The deformation of the bubble in its rebound phase is significantly stronger than in panel (b) of figure 6. As the laser is focused closer to the drop's surface, i.e. 0.3 ≲ Υ ≲ 1.2, the expansion of the bubble provokes the onset of a Rayleigh-Taylor instability. This can be seen in figure 6(d) by the formation of several

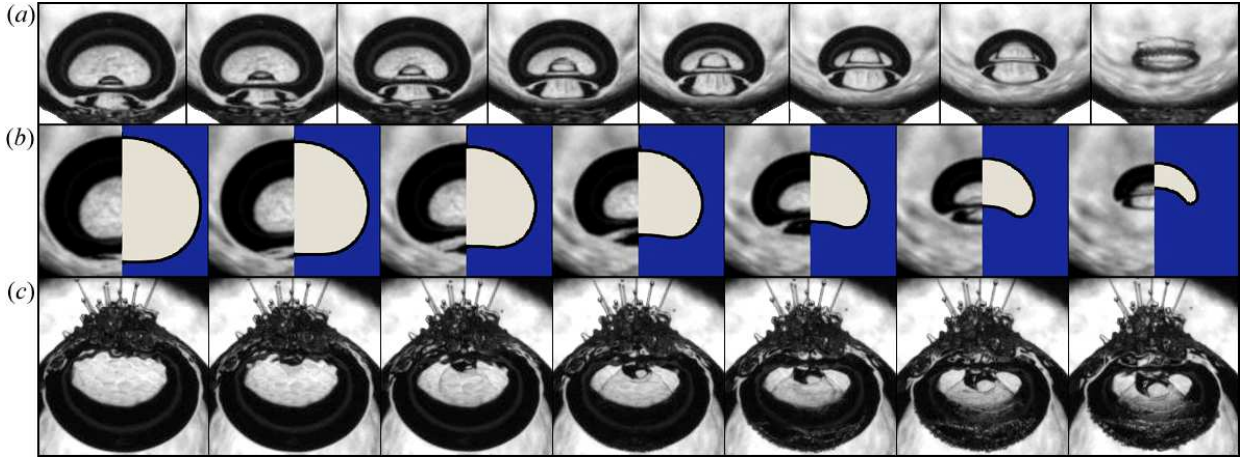

Figure 7: Detailed view of the interior of a jetting bubble. The time between frames is 2 µs. (a) Jet formation for ¥ = 1.6. The frame width is 1.46 mm. (b) Comparison between experimental data and a simulation performed for ¥ = 2.9. The frame width is 1.38 mm. (c) Spray produced by air entering the gas cavity (in which the pressure is lower than the atmospheric pressure) while the jet is formed. The frame width is 2.11 mm.

"spikes" growing from the thin liquid film trapped between the cavity and the surrounding air. At the same time, the bubble collapse (from $t$ = 66 µs) results in an elongated cavity, similarly as in the "bullet jet" case (Rosselló *et al.* 2022). This behaviour is more pronounced for even smaller stand-off distances, as presented in figure 6(e). The dynamics of this particular jet are described in detail in Ref. (Rosselló *et al.* 2022) and correspond to the case where the laser cavity is generated almost directly on the surface of the drop (i.e. $0.01 \lesssim ¥ \lesssim 0.3$). Here, atmospheric gas is trapped after the closure of a conical ventilated splash and later dragged into the liquid by the liquid jet that grows from a stagnation point located on the top of a "water bell" (at the bottom of the frame at $t$ = 36 µs). As a result, an elongated gas cavity is shaped and driven across the drop.

The combined effects of the curved shape of the drop in addition to the diffuse illumination lead to images of the interior of the gas cavity with remarkable clarity. A few examples of this are presented in figure 7.

Panels (a) and (b) of figure 7, reveal the temporal evolution of the liquid indentation into the bubble, as well as the toroidal shape acquired by the gas upon its collapse. Moreover, figure 7(b) demonstrates the accuracy of the numerical simulations to reproduce the jetting process. In panel (c) we see how a perforation of the thin liquid sheet between the cavity and the atmosphere resulted in a spray of aerosol droplets ejected into the cavity during jetting. This event can be explained by the lower pressure inside the bubble compared to the atmospheric pressure and the disruption of the liquid on the upper side of the drop caused by the RTI. The spray front spreads into the cavity and collide with the lower wall of the bubble, disrupting the smoothness of the interface.

The bullet jet case of figure 6(e) distinguishes itself from the other cases by its unique features, i.e. its enhanced shape stability during its formation from an open splash, but also by the near robustness against the surrounding fluid and geometry. Bullet jets have been observed in shallow waters (Rosselló & Ohl 2022) and near flexible or rigid materials, without these conditions affecting their dynamics. Furthermore, in a previous work (Rosselló *et al.* 2022), we demonstrated that the bullet jet is scalable and independent of the orientation of the surface with respect to gravity. In figure 8, we expand the list of remarkable robustness by showing it to exist of various sizes even within a highly curved and finite volume. Here, the bullet jet size was characterised by the ratio between the radius of the initial water bell at its base ($R_{wb}$) and the drop radius $R_d$.

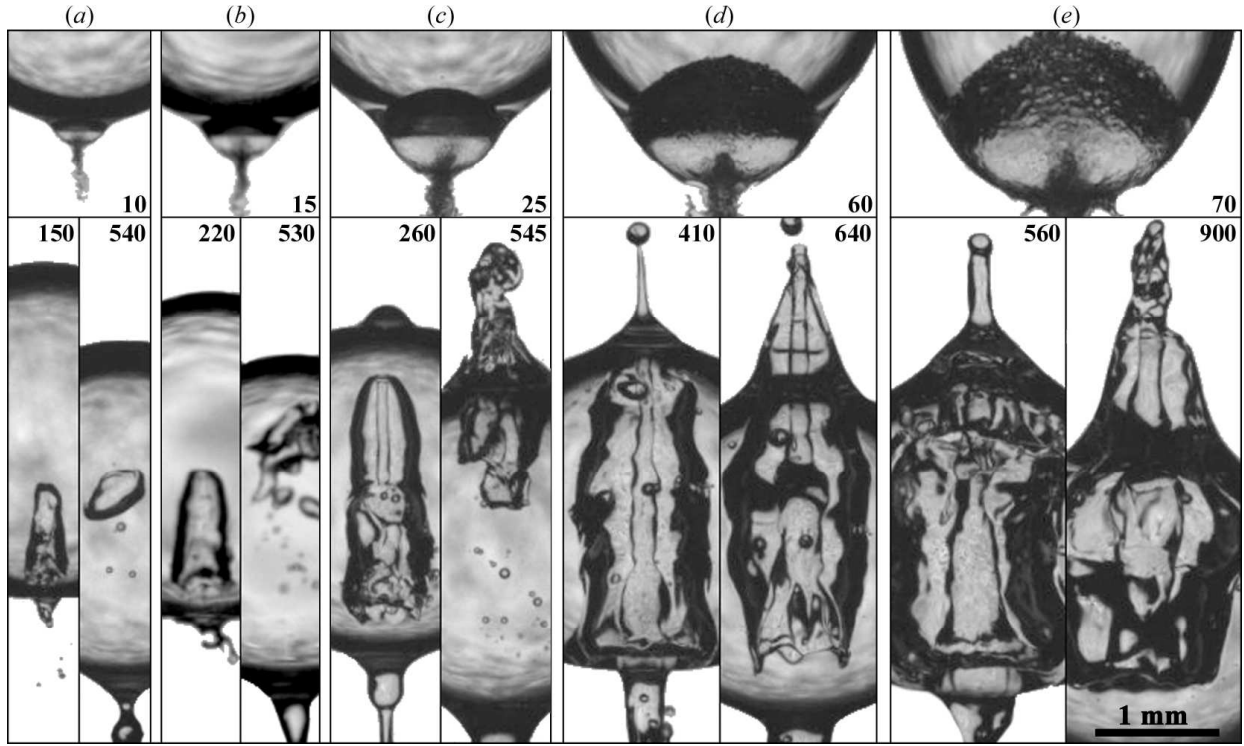

Figure 8: Scalability of the bullet jet in a millimetric droplet. The measurements, organised in columns, show bullet jets formed from different splash sizes. In each column, the upper frame shows the time at which the water bell closes. In the lower frame, composed of two vertical stripes, the time at which the bullet jet is fully developed is shown on the left, and a frame illustrating the position of the jet tip at an advanced time indicated in μs is shown on the right. (a) $R_{wb}/R_d = 0.16$. (b) $R_{wb}/R_d = 0.27$. (c) $R_{wb}/R_d = 0.37$. (d) $R_{wb}/R_d = 0.56$. (e) $R_{wb}/R_d = 0.74$.

The images depict that the penetration depth of both the gas and the liquid conforming to the bullet jet is proportional to the initial splash size. For instance, in figure 8(a) the jet loses its momentum and stops around the middle of the drop, but it crosses the drop for the larger splashes shown in the panels (c) to (e). Remarkably, in the latter case the bullet jet occupies almost the entire drop while still preserving its characteristic features.

The physics behind the evolution of the bubble jetting cases classified in figure 6 can be further explained with the aid of numerical simulations, as presented in figure 9.

Figure 9(a) depicts a purely radial oscillation of both the gas and liquid, found when the bubble is placed in the centre of the drop (i.e. $\Upsilon \rightarrow \infty$). The simulations shown in panels (b) and (c) of figure 9 were computed using the same $\Upsilon$ measured from the experimental cases displayed in the corresponding panels of figure 6. In general, the agreement between the simulations and the experiments is excellent, even though small variations in the size of the experimental and simulated bubbles show some differences in the specific timing of their oscillation cycle. The resemblance can be seen in some of the morphological features that characterise the dynamics of each type of jet at different stages, like the width of the indentation formed during bubble piercing, the shape of the cavity after the first rebound, and the way in which the second collapse evolves in each case. More details on noteworthy features are provided below in figure 10.

In panels (a) to (c) of figure 9 the bubble is initiated with a much larger pressure than the atmospheric gas outside the drop. This pressure difference, which is constant in all directions, accelerates the liquid between the two gas domains. Since this force is proportional to the pressure gradient, the liquid gets accelerated more strongly between the bubble and the nearest part of the drop surface (where the liquid is thinner), causing the drop to bulge out in that location. Within the first few microseconds of the explosive bubble expansion, the pressure within the bubble decreases rapidly and reaches values much smaller than the atmospheric pressure. Thus, the pressure gradient changes its direction and now accelerates the liquid towards the bubble, which first slows down the cavity's expansion and afterward causes its collapse. In the same way as in the expansion phase, the thinnest part of the liquid experiences the strongest acceleration, which ultimately leads to a liquid jet indenting the bubble from the nearest part of the drop surface.

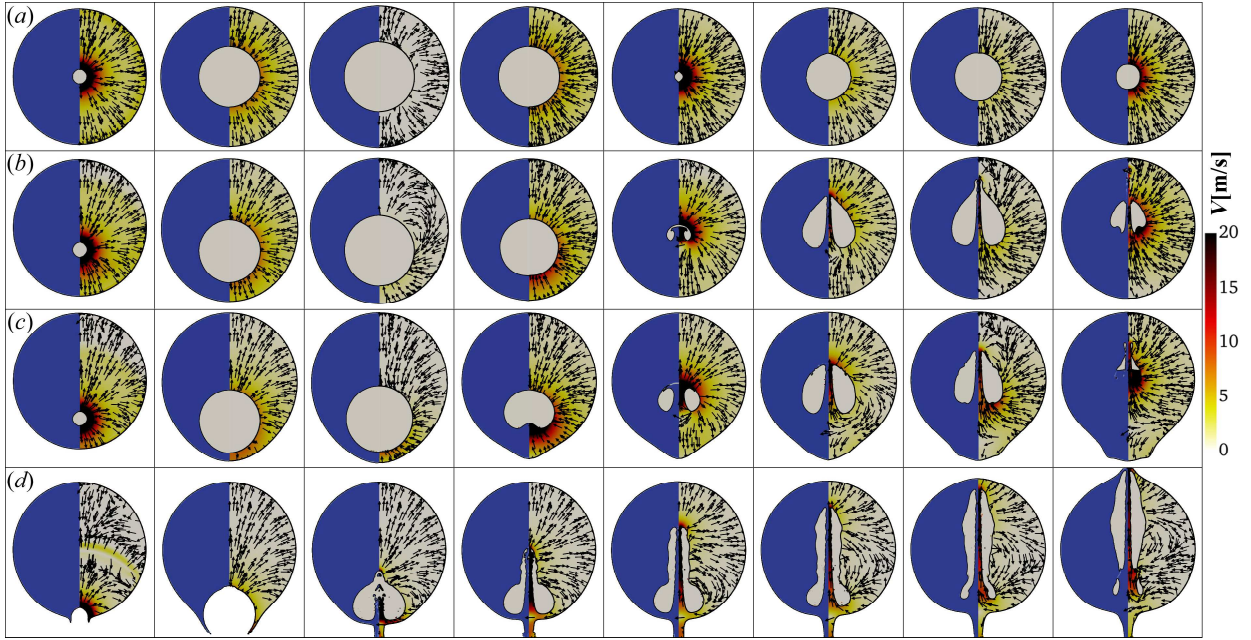

Figure 9: Numerical simulations of the temporal evolution of jets produced inside the drop for different ¥. The simulated drop has a height of 2.88 mm and a width of 2.8 mm as measured in the experiments. The plot shows the gas and liquid phases along with the velocity field. The time between frames is 26 µs for (a)-(c) and 30 µs for (d) starting at $t = 1$ µs in the first frame. (a) Spherical oscillation case, ¥ → ∞. (b) Weak jet case, ¥ = 3.896. (c) Standard jet case, ¥ = 1.518. (d) Bullet jet case, ¥ = 0.028.

The case presented in figure 9(d) differs greatly from the previous cases by the fact that now the bubble is close enough to the drop surface to generate an open cavity, allowing the ejection of the initially pressurised gas inside it into the atmosphere, and later the flow of gas into the expanded cavity before the splash closes again. Once the cavity is closed, it remains with an approximate atmospheric pressure, which prevents it from undergoing a strong collapse as it occurs in the previously discussed cases (a) to (c). The radial sealing of the splash forms an axial jet directed toward the centre of the drop, which pierces the bubble and drags its content through the drop. More details on the mechanisms behind the bullet jet formation can be found in Ref. (Rosselló *et al.* 2022).

As a consequence of the conservation of momentum, the collapse of the gas cavity gives origin to a stagnation point, from which the liquid flows both inside the pierced bubble and away from it in opposite directions. In particular, the stagnation point is not stationary but moves along the axis of symmetry, following a different trajectory in each case. In the case of figure 9(b) the stagnation point shifts towards the surface as the bubble moves deeper into the drop. For the case in figure 9(c) the stagnation point does not reach the surface and its movement is less pronounced. In the bullet jet case, shown in figure 9(d), the stagnation point forms on the apex of the water bell (i.e. the splash after its closure). It then trails the bell's collapse and remains very close to the drop surface afterward, moving slightly towards the drop centre while the bullet jet moves across the drop.

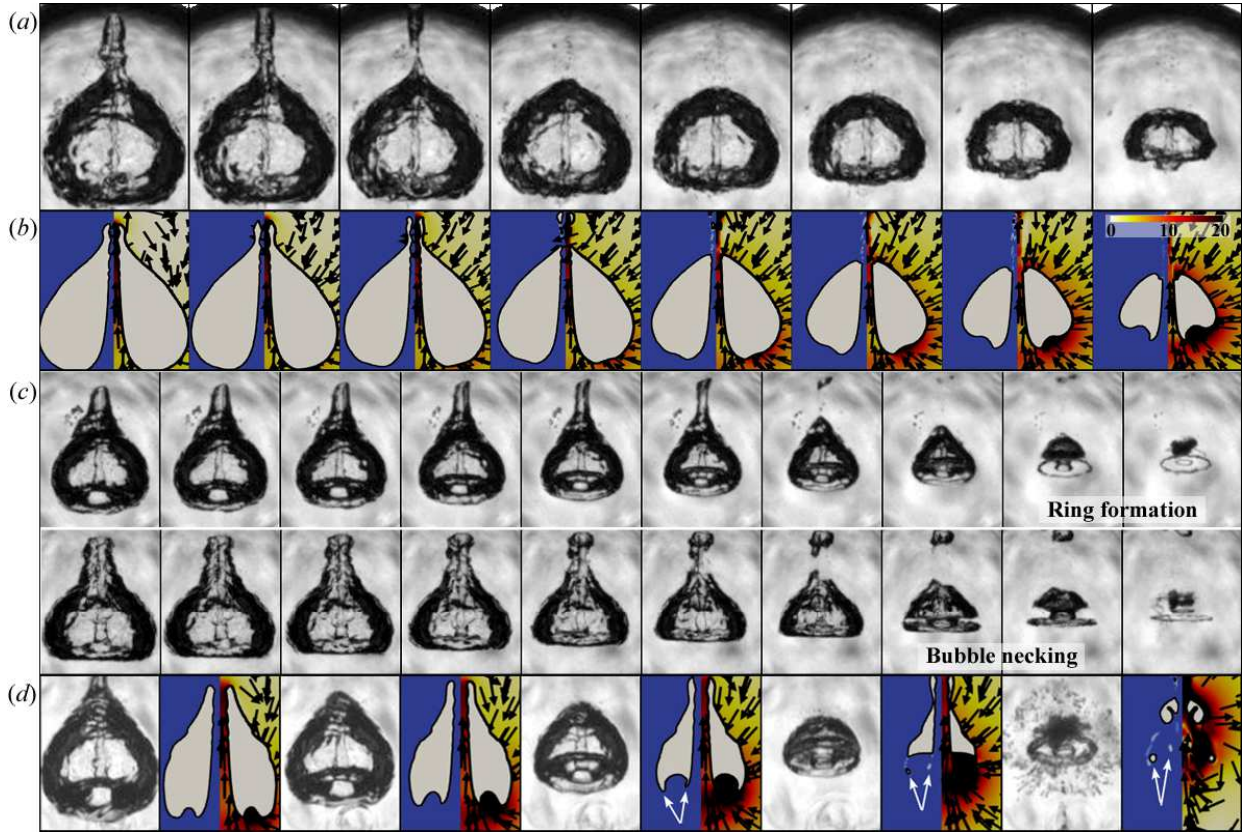

Figure 10: Detailed collapse dynamics of the gas cavity immediately after the jetting of the laser bubble. Experimental (a) and simulated (b) view of the "weak" jet obtained when Υ = 3.9. The images were taken at 200 kfps. (c) Ring formation after the necking of the cavity typically observed on cases with Υ ≈ 1.9. The images were taken at 500 kfps. (d) Direct comparison between experiment and simulation, revealing the precise flow pattern leading to the ring detachment (indicated by the white arrows). The time between frames is 2.5 μs. The colour scale in the simulations corresponds to the one in figure 9.

### 4.2.1. *Cavity dynamics on its second collapse*

After the jetting, the subsequent re-expansions and collapses of the cavities are characterised by the bubble's and the drop's distorted shapes and even more complicated flow fields. A good example of this can be found in the second collapse of the bubbles analysed in figure 10, which shows a significant dependence on Υ.

Figure 10 compares the shape taken by the bubble for two cases with Υ = 3.9 (panels (a) and (b)) and Υ = 1.9 (panels (c) and (d)). Interestingly, the flattened side of the "teardrop" shape acquired by the cavity after the re-expansion develops a curved indentation during its second collapse. The numerical simulations make clear that such an indentation is created by the flow produced by an uneven pressure gradient on the cavity surface. The shape of this ring-shaped indentation visibly changes with Υ. For example, the case presented in figure 10(c) displays an annular bubble necking with the detachment of two gaseous rings as the cavity shrinks. These concentric rings have two different diameters and are arranged in two distinct planes, as highlighted in figure 10(d).

### 4.2.2. *Influence of $R_d$ on the jet dynamics: Behavioural similarity vs. structural similarity*

The bubble dynamics observed in the falling drop case have many similarities with what is typically seen in bubbles collapsing near a planar rigid surface (Lauterborn *et al.* 2018) or a planar free surface (Supponen *et al.* 2016; Rosselló *et al.* 2022). Moreover, the analysis of the values of the stand-off parameter $D^*$ reveals that each type of jet (qualitatively classified according to figure 7 in Ref. (Rosselló *et al.* 2022)) occurs in a comparable range of values of $D^*$. One example of the latter can be found in figure 11.

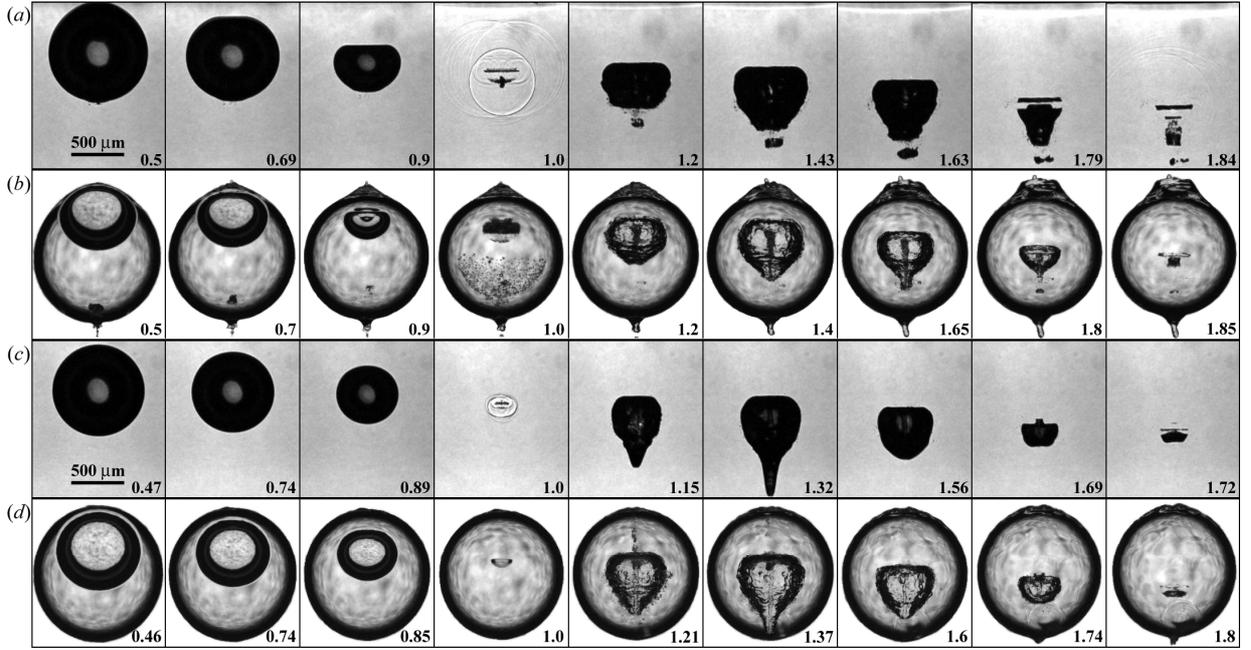

Figure 11: Comparison of cases with similar bubble dynamics and a different curvature of the free surface $R_d$. The panels (a) and (c) show cases where the cavity is produced near a flat free surface. The cases in (b) and (d) show similar bubbles generated inside a drop with a mean radius of 1.42 mm. Here, the numbers represent the time normalised with the time of collapse of the cavities from each case. (a) Here, $D^* = 0.85$. (b) $D^* = 0.88$. (c) $D^* = 1.6$. (d) $D^* = 1.37$.

The parallel found between cases with dissimilar curvature of the liquid surface suggests that, contrarily to the reported observations for bubbles collapsing near concave solid surfaces (Aganin *et al.* 2022), $R_d$ does not have a dominant role in the particular jetting regime adopted by the cavities when the bubbles are located near the free boundary. This statement was confirmed by the numerical simulations depicted in figure 12. There, the dynamics of identical bubbles expanding and collapsing near the surface of the drop, or the flat free surface of an ideally infinite pool, are compared for three stand-off distances $D^*$.

The simulations show that the correspondence between the flat and the curved surface cases is gradually lost when the bubble is placed further away from the drop surface. The deviation between the two cases is already visible in figure 11(c) and (d). There, the jet dynamics are matched only when $D^*$ takes a higher value for the flat free surface measurement. The simulations indicate that this discrepancy starts at around $D^* = 1.2$ (shown in figure 12(c)) and keeps growing for higher values. We can portrait these changes as being enclosed between two extreme scenarios: (1) the bubble is produced right on the liquid surface, generating a bullet jet, which is not affected by the characteristics of the boundaries and thus is independent on $R_d$. As the cavity is placed closer to the drop centre the surface curvature becomes increasingly relevant to the jet dynamics. This is consistent with our definition of $\Upsilon$, since $D^*$ and $\Upsilon$ take similar values for lower values of $d$, and grow apart as the cavity is placed deeper in the drop. (2) When the bubble is almost at the drop centre (i.e. $\Upsilon \to \infty$) there is no jetting for the curved case. However, in the flat surface case the jetting still occurs for comparable values of $D^*$ (e.g., $D^* \sim 2$), demonstrating how the curvature weighting factor $\chi$ becomes increasingly relevant.

It is important to stress that if the bubble is placed near the drop boundary, for instance at $D^* \lesssim 1.4$, the discrepancies found in the jetting dynamics of a bubble in the "semi-infinite" liquid pool when compared with the droplet case are mainly provoked by the surface curvature, and not by the dissimilar extension of the liquid below the gas cavity. This particular point is corroborated in the Appendix A by means of complementary measurements and numerical simulations of jetting bubbles in the proximity of a hemispherical tip of a cylindrical water column.

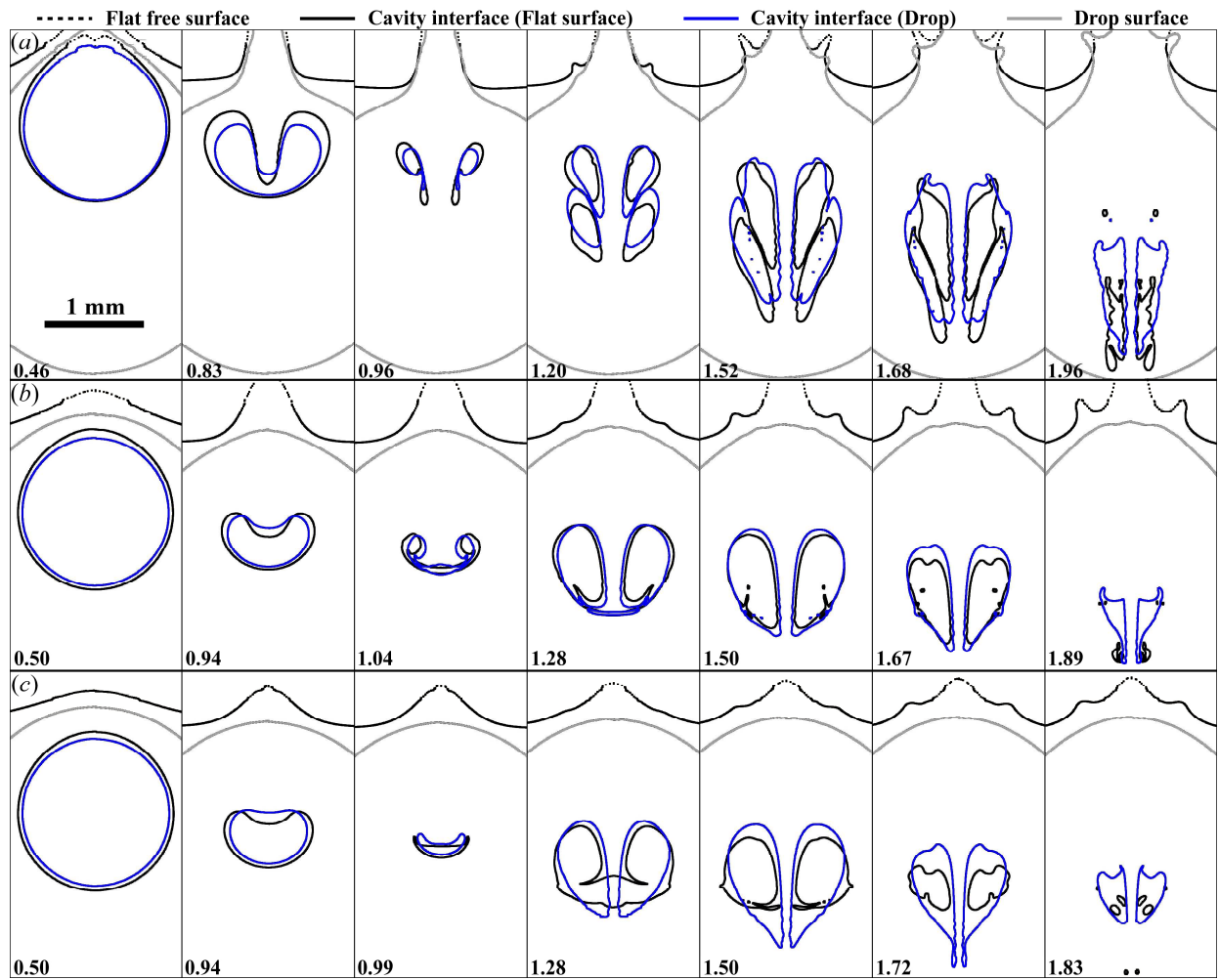

Figure 12: Jetting dynamics of identical bubbles produced near a flat surface or the curved surface of a droplet ($R_d = 1.42\,\mu$m). The non-dimensional time, indicated by the numbers, was normalised with the collapse time of each bubble. (a) Here, $D^* = 0.61$. (b) $D^* = 1.02$. (c) $D^* = 1.23$.

So far, we have classified and compared the characteristics of the jetting regimes produced at different $D^*$ qualitatively, i.e. based on their general morphological features as presented in previous works (Supponen *et al.* 2016; Rosselló *et al.* 2022). In the following, we will refer to this as *behavioural similarity*. An alternative and more precise way of analysing the spatial correlation between the dynamics of two different jets can be achieved by contrasting the pixel distribution on the video frames to find common features between images. This quantitative comparison method is usually referred as *structural similarity* analysis and can be implemented using different image scanning algorithms (Sampat *et al.* 2009). Here, we use the *complex wavelet structural similarity index* (CW-SSIM) (Zhou & Simoncelli 2005) to evaluate the correlation between the temporal evolution of two different jetting cavities. The CW-SSIM approach has some advantages over direct pixel to pixel comparison methods (e.g. intensity-based) or the simpler versions of the structural similarity index (e.g. SSIM). For instance, it accounts (up to some point) for both intensity variations and non-structural geometric distortions like object translation, scaling and rotation (Sampat *et al.* 2009). The CW-SSIM index can take values ranging from zero (if there is no correlation at all) to one (when the images are identical).

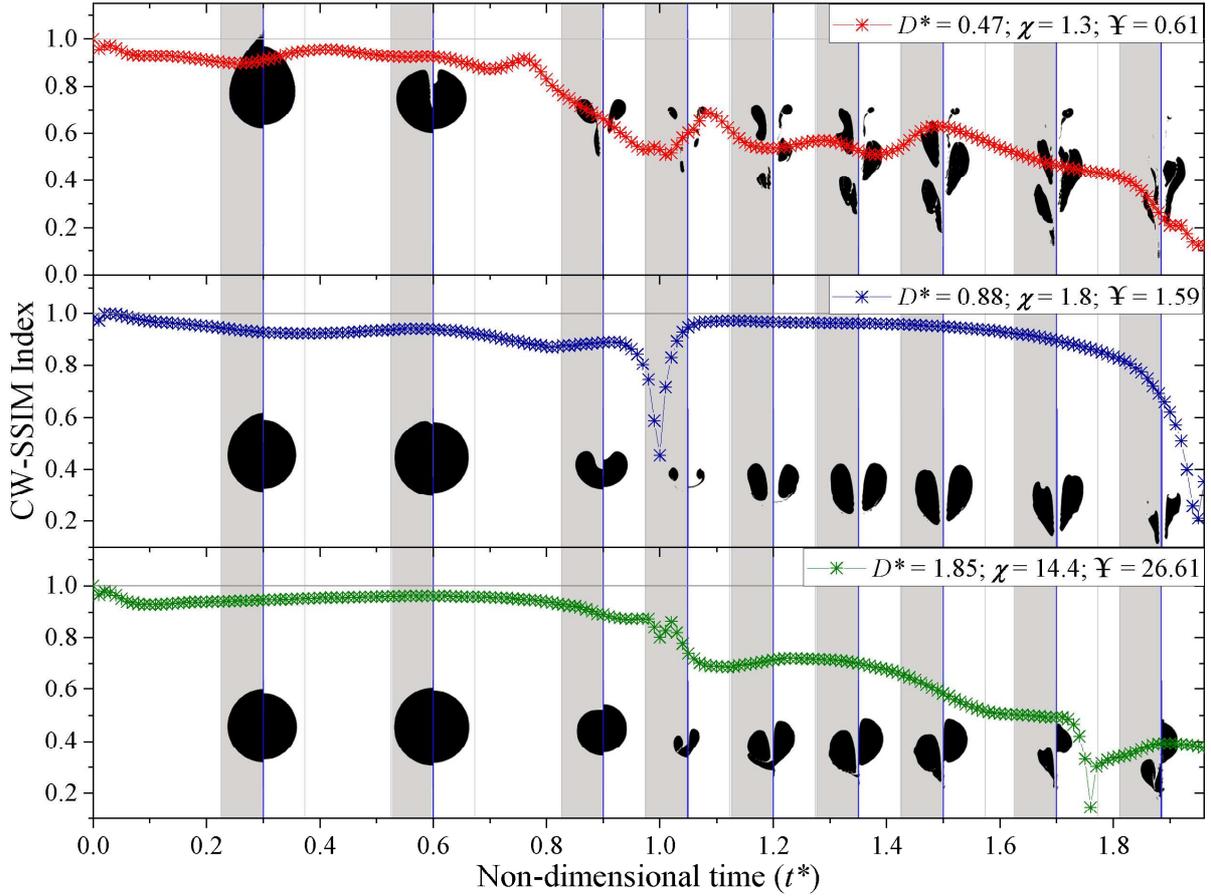

Figure 13: Structural similarity of the dynamics of jetting bubbles produced near a flat surface or the drop boundary at an identical stand-off distance. Here, the temporal evolution of the CW-SSIM index is presented for three examples corresponding to cases with $D^* = 0.47$, $D^* = 0.88$, and $D^* = 1.85$. The insets show a comparison of the images of both simulated bubbles. The frames are centered at specific non-dimensional times (blue vertical lines), displaying a half frame corresponding to the flat surface case (grey background on the left) and a half frame taken from the drop case with the same $D^*$.

In figure 13, we contrast the dynamics of two bubbles initially located at a distance $D^*$ from a flat or a curved surface as already done in figure 12, but this time using the CW-SSIM index to evaluate their similarity. The non-dimensional time ($t^*$) was computed using the collapse time of the bubbles for each case. Figure 13 presents plots of the temporal evolution of the similarity index next to a series of selected frames at $t^* = 0.3; 0.6; 0.9; 1.05; 1.2; 1.35; 1.5; 1.7;$ and 1.9, which illustrate and compare the shape of the cavities in the flat boundary case (grey background on the left side) and the drop case.

The results expose the differences between the behavioural similarity and the structural similarity approaches, i.e. two bubbles can have the same jetting regime but still have dissimilar structures. This is observed in bubbles at lower stand-off distances like the case with $D^* = 0.47$ in figure 13. The discrepancy can be explained by the higher degree of fragmentation of the bubble after jetting, acquiring an elongated shape in regimes with a ventilated cavity, or where the liquid layer between the gas in the bubble and the atmosphere is affected by the RTI. In particular, the cases producing an open cavity (i.e. $D^* \lesssim 0.35$) were not suitable for the structural similarity analysis. Here, the fluctuations of the splashing dynamics observed in the numerical simulations and the impossibility to define a collapse time due to the non-collapsing nature of those cavities (see figure 6(d) and (e)) prevented us to perform a reliable assessment of the CW-SSIM index.

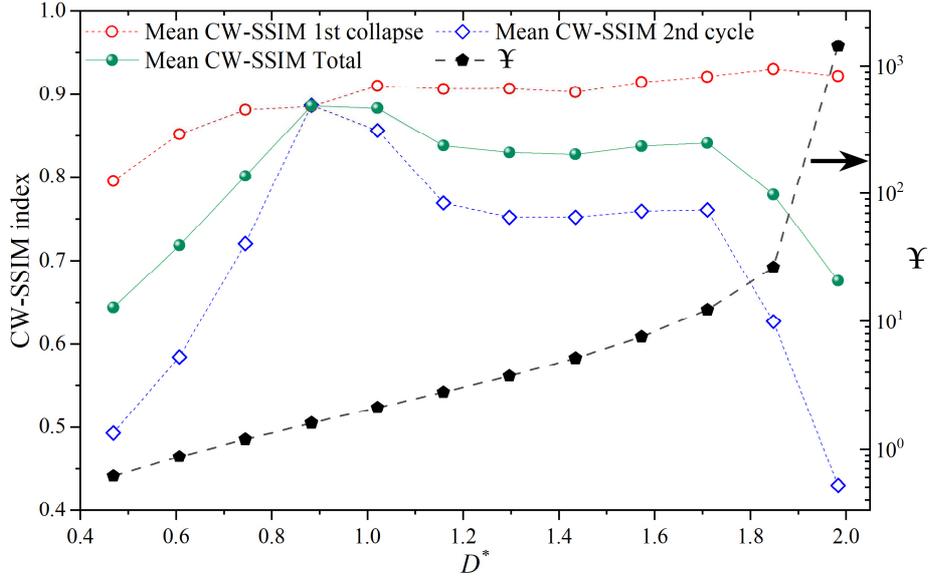

Figure 14: Structural similarity between cases with bubbles seeded at different stand-off distances from a flat free surface or inside the drop. The curve shows the mean value of the "total" CW-SSIM index (green) computed as the average of the mean indices observes in the first bubble collapse (red) and in its second oscillation cycle (blue). The sudden increase in $\Upsilon$ (black markers) as we seed the bubble close to the drop centre is linked to a decay in the similarity between the cavities.

As the laser bubble is produced deeper into the liquid, both structural and behavioural approaches lead to the same conclusions (previously discussed in figure 12). The similarity found in the development of bubbles near surfaces with or without curvature is excellent for some stand-off distances, e.g. around the middle point located between the liquid surface and the drop centre. One example of this is shown in the central panel of figure 13 corresponding to $D^* = 0.88$. As we already mentioned, near the centre of the drop (i.e., $D^* \simeq 2$) the difference in the anisotropy in both cases produces dissimilar bubble oscillations (see the lower panel of figure 13).

A more general overview of those three scenarios is presented in figure 14, where the mean value of the CW-SSIM index is plotted along with $D^*$ and $\Upsilon$. Considering that all the bubbles have a very similar initial expansion phase, only the times corresponding to the first collapse and the complete second oscillation cycle were computed in the mean value of CW-SSIM. After the second collapse, the bubble is heavily fragmented and there is no longer a recognisable structure. Figure 14 confirms that the structural similarity is rather poor for bubbles near the surface (i.e. $0 \lesssim D^* \lesssim 0.7$). Around $D^* = 0.9$, the similarity index reaches a peak where the match is excellent. A good agreement is sustained over a range of $D^*$ values between approximately 0.7 and 1.7, meaning that even when the features of the cavities are not identical they have a similar distribution of the gas phase (and the same jetting regime). For $D^* \gtrsim 1.7$, the similarity index suffers an abrupt fall and the value of $\Upsilon$ diverges as the bubble seeding position gets closer to the drop centre. This is consistent with the definition of $\Upsilon$, which relates its magnitude to the influence of the drop curvature in the bubble dynamics.

The experimental results displayed in figure 11 suggest that there is a correspondence between the dynamics of a bubble seeded with a given $D^*_{\text{flat}}$ in the flat surface case and the temporal evolution of a bubble with $D^*_{\text{drop}}$ inside the droplet. We explore this apparent

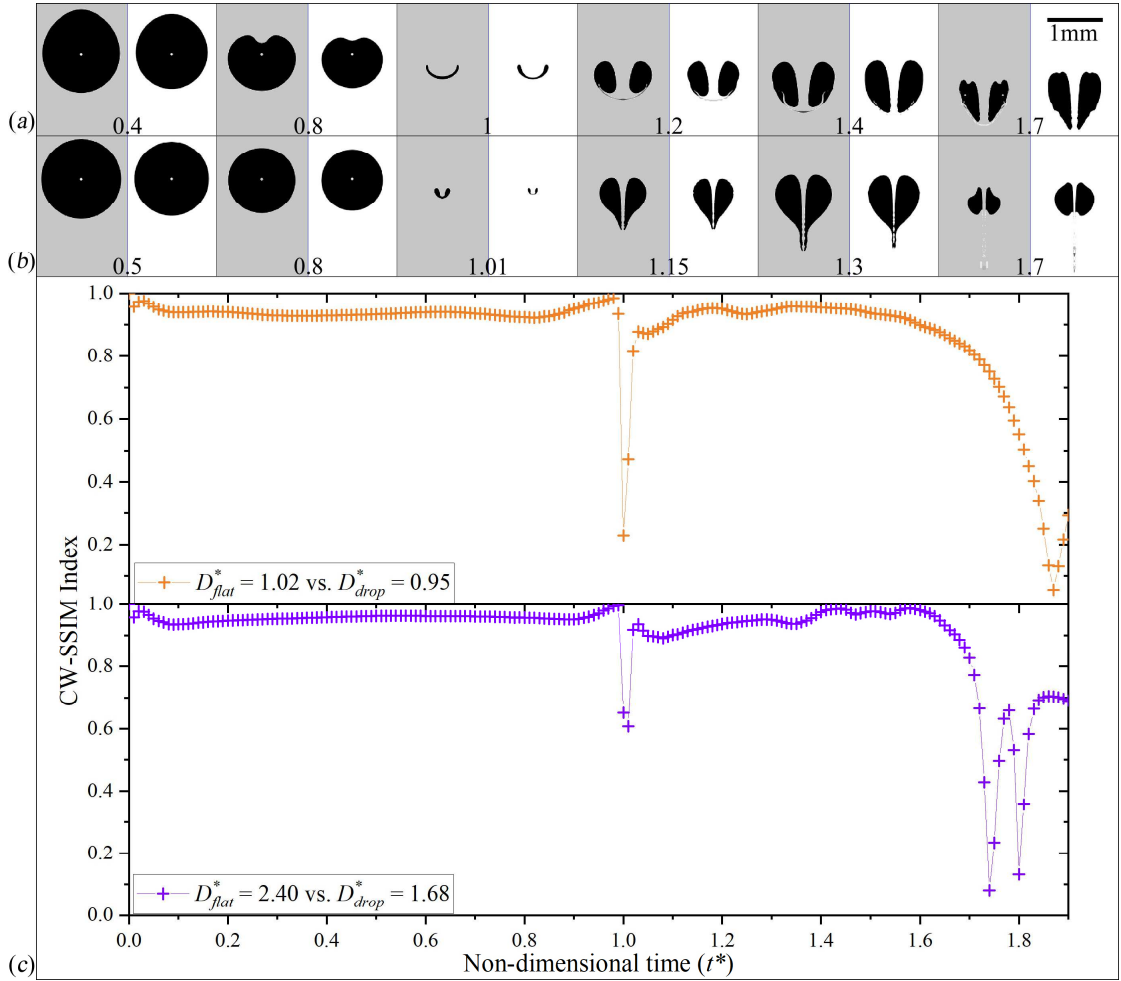

Figure 15: Similarity study of the jetting dynamics of bubbles near a flat surface (grey background) or a drop (white background) with a different $D^*$. The initial position of the bubbles was matched by performing a vertical shift on the drop case simulations. For each value of $D^*_{\text{flat}}$ there is one value of $D^*_{\text{drop}}$ with similar dynamics, i.e. producing the maximum CW-SSIM index when a simulation of a given $D^*_{\text{flat}}$ is compared against simulations with every possible value of $D^*_{\text{drop}}$. The numbers indicate non-dimensional time $t^*$. (a) For $D^*_{\text{flat}} = 1.02$ the maximum average CW-SSIM index was achieved with $D^*_{drop} = 0.95$. (b) The best match for $D^*_{\text{flat}} = 2.40$ was $D^*_{\text{drop}} = 1.68$. (c) Temporal evolution of the CW-SSIM index for the cases on panels (a) and (b).

"equivalence" between values of $D^*_{\text{flat}}$ and $D^*_{\text{drop}}$ in figure 15 and figure 16. Since the result of the CW-SSIM analysis is affected by a significant translation of the objects being compared, the initial positions of the bubbles were matched by performing a vertical shift on the drop case simulations. Figure 15 shows evidence of the mentioned correspondence by presenting two examples, one with $D^*_{\text{flat}} = 1.02$ and $D^*_{\text{drop}} = 0.95$, and a second one where the bubble is closer to the drop centre, i.e., $D^*_{\text{flat}} = 2.40$ and $D^*_{\text{drop}} = 1.68$.

These two examples in figure 15 prove that there are pairs of $D^*$ values where the similarity between the dynamics of bubbles produced near two surfaces with uneven curvature is remarkable, at least during the whole period comprised in the first two oscillation cycles. This correlation analysis was performed for an extended range of values of $D^*$ to find that for each value of $D^*_{\text{drop}}$ there is one value of $D^*_{\text{flat}}$ with similar dynamics, i.e. which maximise the CW-SSIM index when a simulation made with that particular $D^*_{\text{flat}}$ is compared against simulations with every possible value of $D^*_{\text{drop}}$. As shown in figure 16, the dependence of the equivalent stand-off distance starts as a linear function in the proximity of the surface and grows rapidly as the bubble is placed deeper in the liquid. Interestingly, the linear fit performed on smaller values of $D^*$, which meet the conditions for the CW-SSIM analysis, projects a ratio between equivalent $D^*_{\text{drop}}$ and $D^*_{\text{flat}}$ near 1 when the seeding position approaches the surface. Now, bearing in mind the definition of $\Upsilon = \chi\, D^*$, the previous observation is consistent with the limiting case at the surface where $\chi \to 1$, meaning that the curvature does not play a significant role for the bubble jetting dynamics. At the other extreme, i.e. as $d \to R_d$, $\Upsilon$ diverges, indicating a strong influence of the drop geometry on the bubble evolution. At this point, it is important to stress that in this context, "equivalence" does not mean that the dynamics are identical, but their structure is "as similar as it could be" for the matching of $D^*_{\text{flat}}$ and $D^*_{\text{drop}}$. The same clarification applies to bubbles with the same $\Upsilon$, which is a multivalued function as explained in section 2.1.

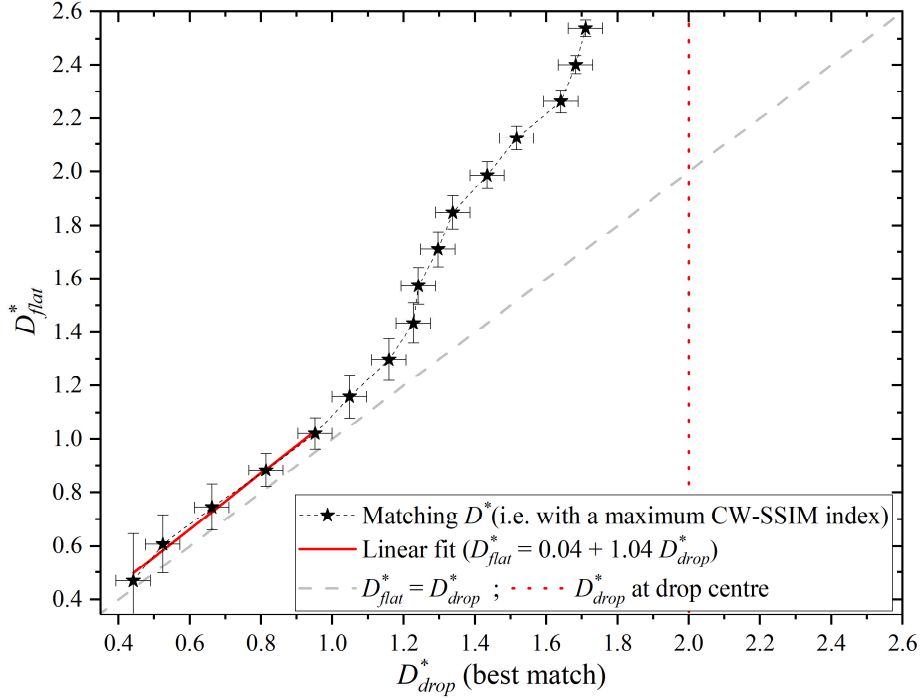

Figure 16: Best match between the dynamics of bubbles in the flat boundary and the drop cases for different stand-off distances, i.e. $D^*_{\text{flat}}$ and $D^*_{\text{drop}}$. For each value of $D^*_{\text{flat}}$, the best match was obtained by finding the corresponding value of $D^*_{\text{drop}}$ that maximises the mean CW-SSIM index. As indicated by the parameters of the linear fit, the best match is found at similar values of $D^*$ at the lower depths, but they become increasingly different as $d \to R_d$, where the bubble is seeded at the drop centre (indicated with a vertical dotted line). The dashed grey line was added as a visual reference. The vertical error bars represent the deviation of CW-SSIM from the perfect similarity case (i.e. CW-SSIM = 1)

### 4.3. *Radial bubble oscillations*

In the previous section, we studied features found in the dynamics of an axisymmetric jetting bubble. Let us now have a closer look at the only case with spherical symmetry, i.e. where the laser cavity is placed in the centre of the drop ($\Upsilon \to \infty$). In this scenario, the bubble undergoes several spherical oscillations with a decaying amplitude, as commonly observed in laser bubbles created in unbounded liquids (Liang *et al.* 2022). Figure 17(a) presents a comparison between an experiment and simulated data computed using the VoF solver, finding an excellent agreement. Like for the previously simulated results, here we applied the correction script that accounts for the distortion induced by the drop curvature.

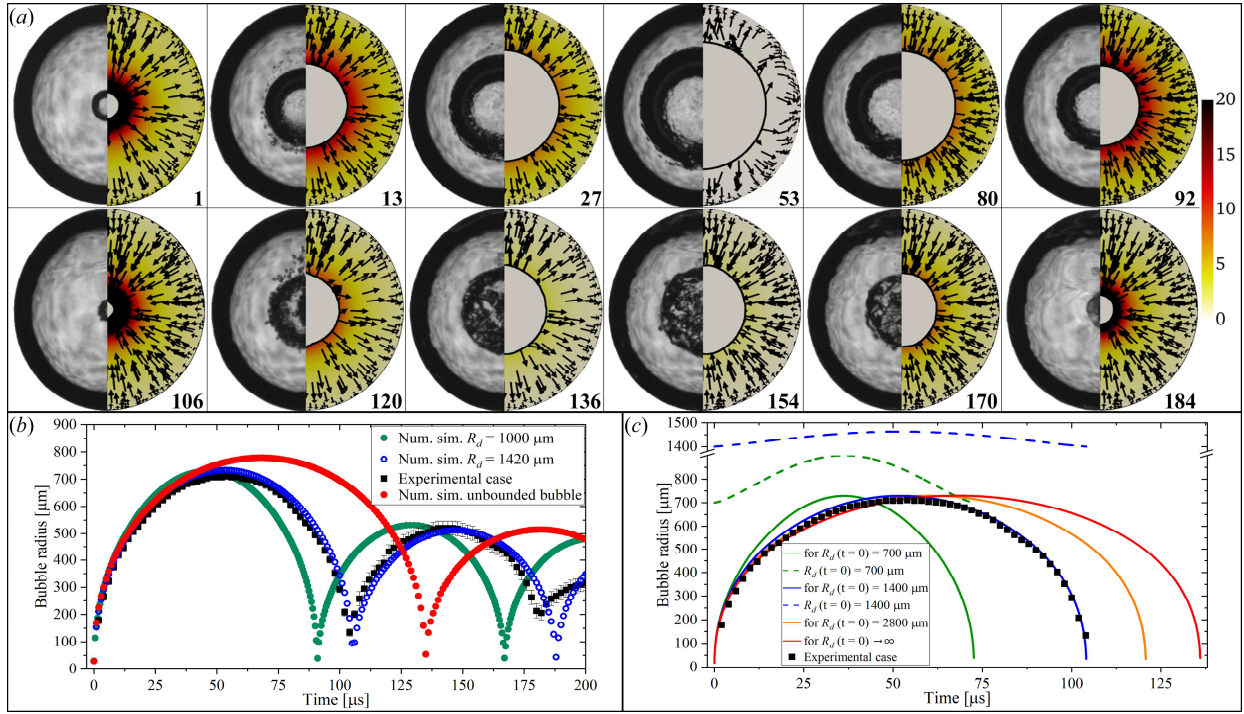

Figure 17: Direct comparison between the experiment and a numerical simulation for a case where the laser bubble is placed at the centre of the drop. (a) The median diameter of the drop is 1.42 mm. The simulated images showing the velocity field have been remapped to account for the distortion provoked by the curvature of the drop. The numbers indicate time in μs and the colour scale is given in m/s. (b) Radial dynamics of the experimental and simulated bubbles. The experimental radius was obtained by fitting a circle on the bubble. The radius in the simulations was estimated using the gas volume (i.e. the spherical equivalent radius). The results were compared with the unbounded case to find that the bubbles inside the drop have a shorter expansion/collapse cycle. (c) Bubble dynamics is obtained with a modified Rayleigh-Plesset model for different drop sizes. The radii of the larger drops remain almost unaltered during the bubble oscillation.

Figure 17(b) depicts the temporal evolution of the bubble radius $R(t)$ for the examples in panel (a). In addition, it presents $R(t)$ calculated for a case of a drop of an ideally infinite size, which corresponds to the case of an unbounded liquid domain. The initial conditions in the VoF model were chosen to match the experimental $R^*_{max}$, and then the other cases were simulated maintaining the same parameters while changing the drop size. Notably, the bubble computed with CFD reaches a slightly larger maximum radius as the liquid layer thickness is increased to infinity (i.e. an unbounded bubble case) and thus also has a larger collapse time. This might be due to the effect produced by the consecutive (and alternating) tension and pressure waves interacting with the bubble during its expansion (see figure 17(a)).

To shed some light on this matter, we use a spherical bubble model based on a modified Rayleigh-Plesset model (RP) (Obreschkow *et al.* 2006; Zeng *et al.* 2018) that accounts for the finite droplet size, viscosity of the liquid and interfacial tension. It is worth noting that in those previous works the millimetre sized droplet was sitting on the top of a blunt needle or deformed into an ellipsoidal shape by a strong levitating acoustic field. This lead to a non-spherical boundary conditions that affect the bubble dynamics. In the present analysis, the droplet is nearly perfectly spherical, thus matching with the purely spherical RP model within a droplet. The results, presented in figure 17(c), show that the bubble grows up to almost the same size independently of the drop size. For the same initial conditions set in the VoF model ($p_g(t = 0) = 1.69$ GPa and $R_b(t = 0) = 17.3$ μm) the work is almost completely done against the surrounding pressure ($p_\infty = 1$ bar) while surface energy and viscous dissipation is negligible. Yet, for smaller droplet volumes, the inertia is reduced and therefore the expansion time to maximum bubble radius and the almost symmetrical collapse reduce, too.

For the particular initial conditions, both models agree on the elongation of the oscillation cycle, however they predict dissimilar results on the maximum radius reached by the bubbles. The simple Rayleigh-Plesset model is used to give a comparison to the VoF simulations and to evaluate the impact of the shock wave (and its reflections) on the bubble dynamics. From figure 17(c) we can infer that the maximum expansion of the bubble is nearly independent of the droplet size, while in the VoF simulations it is not. The VoF model accounts for the reflected wave, thus the discrepancy suggests that upon the acoustic soft reflection of the shock wave momentum is imparted on the droplet interface. The importance of reflected waves on cavitation nucleation in confined liquid samples was recently also found for an acoustic hard reflection where the bubble expansion was lowered (Bao *et al.* 2023). A more comprehensive formulation for the bubble dynamics than the RP model which incorporates both the bubble-shockwave interaction and compressibility effects can be found in Zhang *et al.* (2023).

### 4.4. *Drop surface instabilities*

In the previous sections, the formation of radial liquid jets growing from the drop surface in the shape of "spikes" was mentioned. As explained above, this phenomenon stems from an initial perturbation of the liquid interface and the posterior ejection of liquid produced by the Rayleigh-Taylor instability. This kind of instability occurs when the rapid expansion or the collapse of the bubble wall accelerates a thin liquid layer trapped between the cavity and the atmospheric gas, producing a pattern of ripples on the drop surface that grow further in the consecutive bubble oscillations. A clear example of the events leading to the onset of this kind of instability in this particular experiment is shown in figure 18(a). There, the acoustic emissions from the laser dielectric breakdown nucleate a cloud of bubbles within the drop. As the cavity expands, all these smaller bubbles are incorporated (by coalescence) into the main bubble, producing a series of dimples on the bubble surface, visible at $t = 50\,\mu s$ of figure 18(a). These dimples may contribute to the later destabilisation of the drop surface, which is highly dependent on the ratio $R_{max}^*/R_d$. Additionally, $R_{max}^*/R_d$ determines the liquid layer thickness and its acceleration by the bubble/drop dynamics. The ripples in the drop surface become noticeable just after the first bubble collapse (i.e. $t = 130\,\mu s$) and grow significantly during the bubble re-expansion, as shown at $t = 190\,\mu s$. However, the most dramatic events take place after the second bubble collapse (i.e. at $t = 230\,\mu s$). There, the ripples grow into liquid "spikes" which lead to the detachment of small droplets due to the action of the Rayleigh-Plateau instability, as shown in figure 18(b). At the same time, the second bubble collapse releases a strong shock wave in the radial direction. This shock wave interacts with the array of meniscus-shaped pits on the liquid surface to produce fast radial jets (see the frame at $t = 420\,\mu s$). The later sequence is clearly captured in the frames of figure 18(c). It is important to note that this complex phenomenon not only depends on the shock wave strength but also requires certain conditions to be met (Tagawa *et al.* 2012; Peters *et al.* 2013), like a minimum depth and curvature of the pits, which may explain the absence of "spikes" during the first bubble collapse.

To further analyse the onset of these instabilities, we varied the energy of the laser pulse, hence producing bubbles with various sizes and thus with distinct ratios $R_{max}^*/R_d$. The results are presented in figure 19. Even when the extreme image distortion produced near the drop interface prevent us to obtain an accurate value of the bubble radius, these measurements make evident that the amplitude of the ripples increases with increasing $R_{max}^*$ and with each consecutive bubble oscillation.

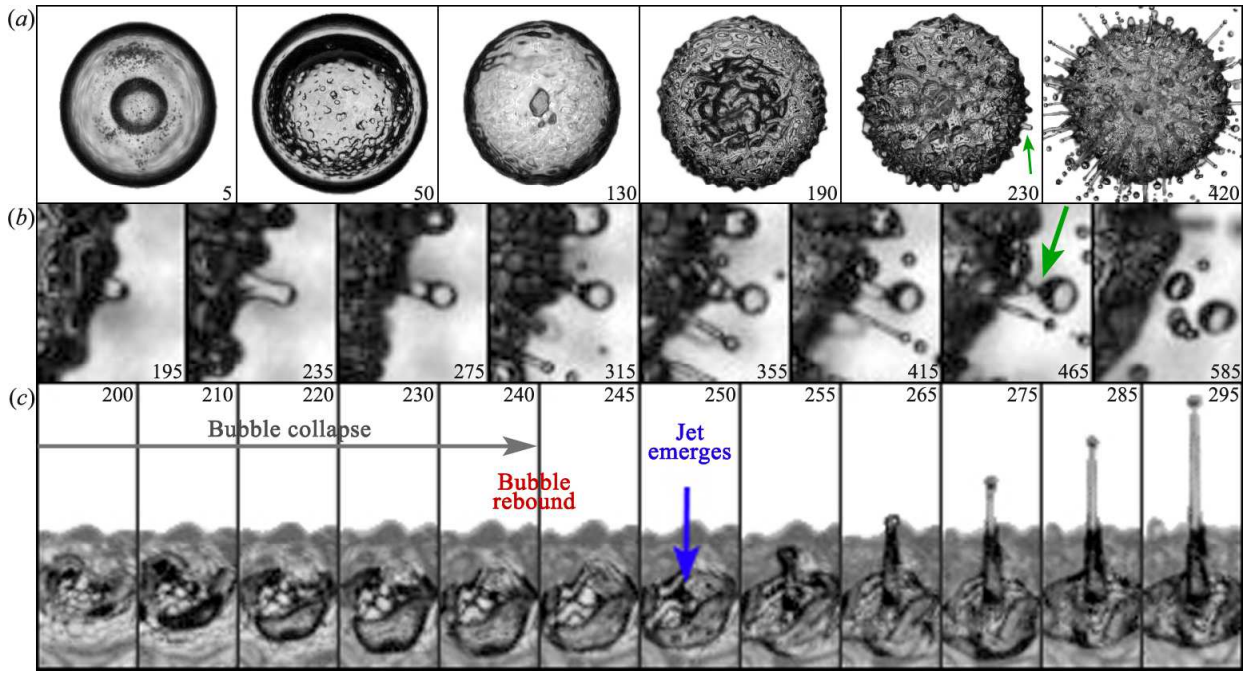

Figure 18: Drop surface destabilisation mechanisms. The mean drop radius is 1.42 mm and the numbers represent time in µs. (a) As the main bubble expands, the secondary, acoustic cavitation bubbles produce small dimples on the gas cavity surface (e.g. at 50 µs). Those may promote the formation of a series of ripples during the bubble collapse (at 130 µs). As the bubble re-expands, the Rayleigh-Taylor instability causes the growth of liquid "spikes" that later lead to the detachment of small drops due to the Rayleigh-Plateau instability, as indicated with a green arrow in panel (b). There, the frame width is 570 µm. At the same time, the second collapse of the bubble enhances the surface irregularities and pits that appear in the areas between the ripples. The shock wave emitted during the second collapse gives origin to fast liquid jets ejected from the centre of the pits, as highlighted with a blue arrow in panel (c). The frame width in this sequence is 490 µm. The full video is available in the online supplementary movie 9.

In figure 19(a) the expansion of the bubble is not sufficient to visibly disturb the drop's spherical surface. In the case shown in figure 19(b) the bubble's first collapse does not break up the drop surface, however, a mild wave pattern is observed on the surface after the bubble re-expansion (at $t = 420$ µs). In spite of the presence of these low amplitude ripples, no radial jets are ejected from the drop upon the second bubble collapse. When the ratio $R^*_{max}/R_d$ is further increased, as shown in figure 19(c), we find very similar dynamics of the bubble/drop system, but now the valleys between the ripples (and the acoustic pressure wave) are deep enough to trigger the radial jetting. This confirms the existence of threshold conditions for the "spikes" to be formed. In the remaining cases presented in panels (d) to (f) of figure 19, the general dynamics of the bubble/drop system are very similar to the previous cases, although as the laser pulse energy is increased the instabilities become perceivable at an earlier time. For example, in figure 19(f) liquid "spikes" are already formed after the first bubble collapse (Zeng *et al.* 2018).

Figure 20 shows VoF simulations of the Rayleigh-Taylor instability found on the drop surface. From figure 20(a) it is clear that the instability is grown by the volumetric oscillation of the bubble, while the shock waves emitted from the bubble upon its creation (and later at its collapse) accelerates the ripples on the drop surface and form the thin "spikes". This kind of simulations was previously performed by Zeng *et al.* (2018) for an ellipsoidal droplet with the RTI manifesting only in a reduced region of the surface located on the drop poles. In the present work, we study a nearly spherically symmetric case where the spikes have no preferred origin, i.e. they escape the droplet isotropically.

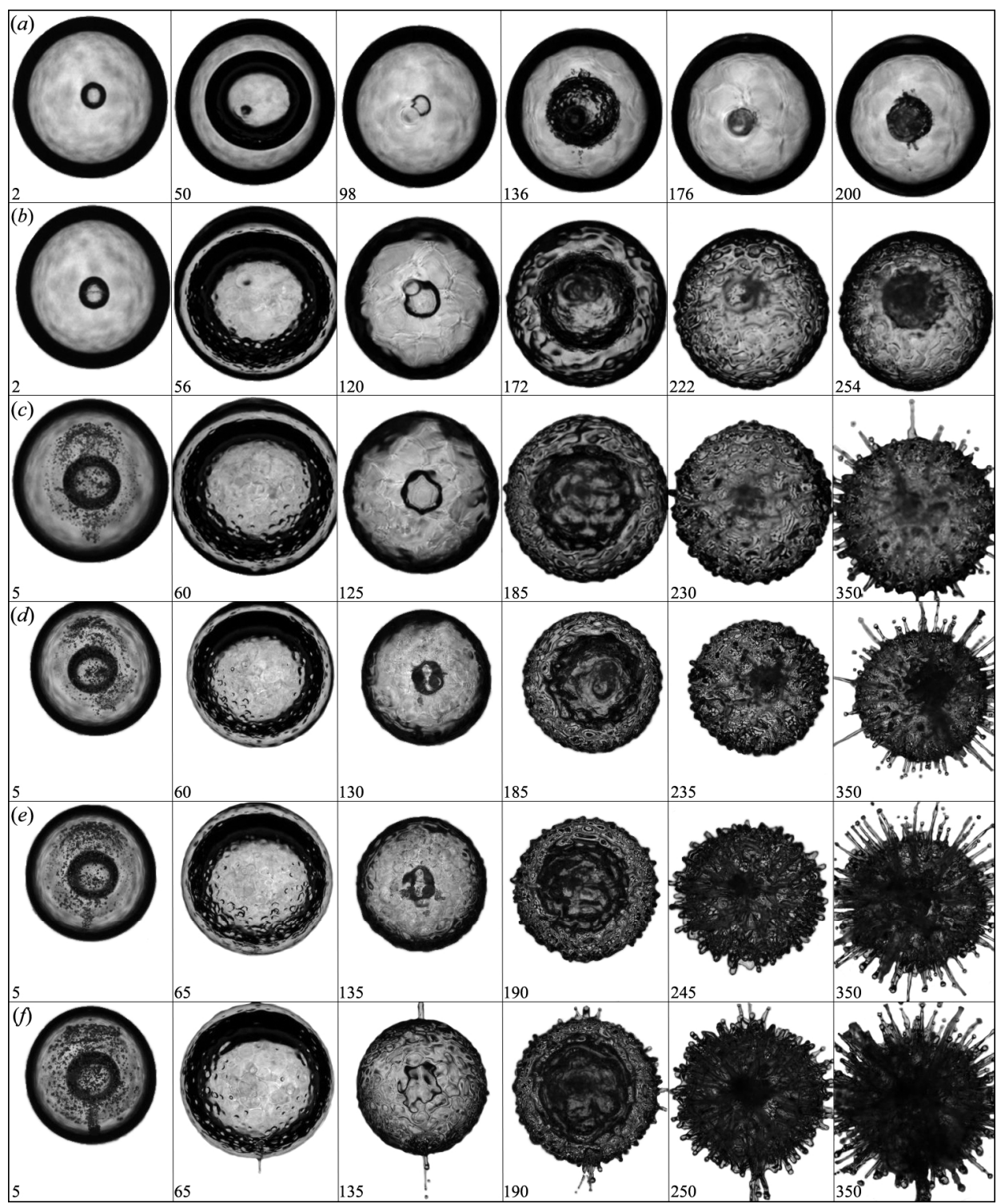

Figure 19: Onset of the drop surface instabilities for bubbles produced with different laser pulse energies. The mean drop radius is 1.42 mm and the numbers represent time in μs. (a) Here, the energy of the laser pulse is $L = 1.9$ mJ. (b) $L = 3.1$ mJ. (c) $L = 3.9$ mJ. For this energy, the RTI affects the drop surface enough to produce liquid ejection after the second bubble collapse. (d) $L = 4.6$ mJ. (e) $L = 5.2$ mJ. (f) $L = 6.4$ mJ. Note that panels (d)-(f) are shown in wider frames than the panels (a)-(c) to show the larger "spikes". A full video of panel (f) is available in the online supplementary movie 10.

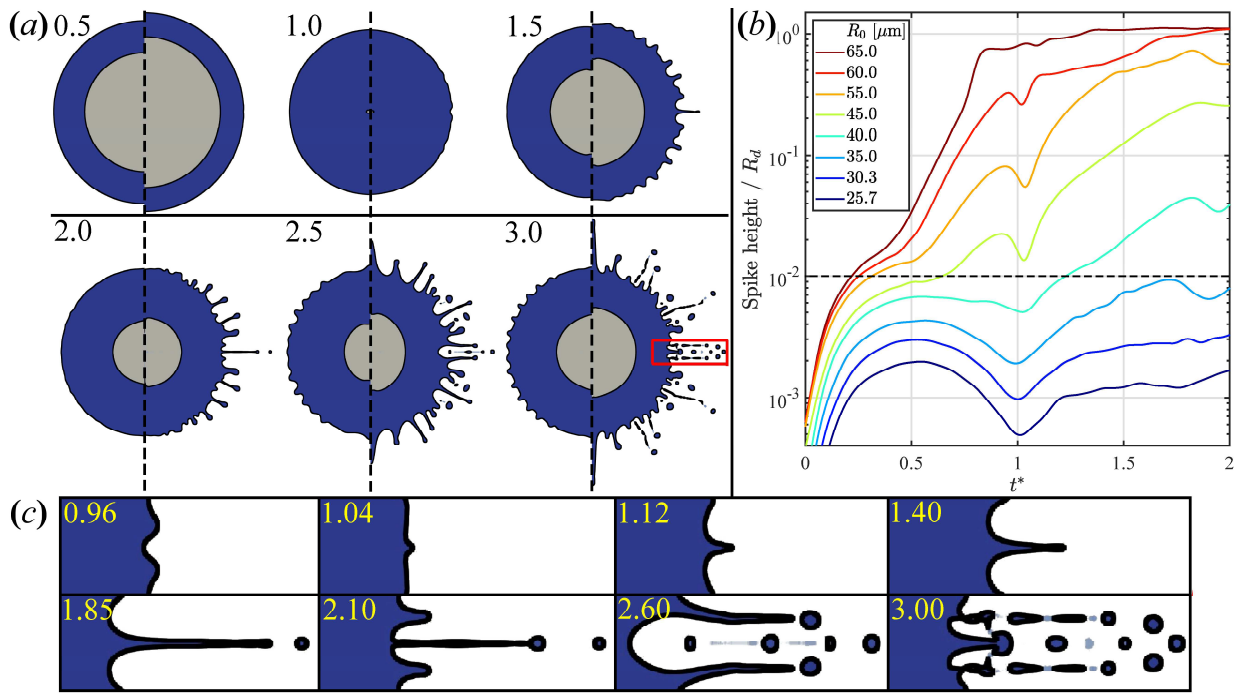

Figure 20: Numerical simulations of the instabilities development at the surface of the drop. (a) Selected frames for a case with $R_0 = 35\,\mu$m (left) and $R_0 = 45\,\mu$m (right). The non-dimensional time $t^*$ is shown on the top-left of each frame. (b) Temporal evolution of the Rayleigh-Taylor instability (RTI) spike height for various $R_0$. An *ad hoc* threshold for the instability onset is indicated by a dashed line at 1 % of $R_d$. (c) Selected frames of a zoomed view of the drop surface for $R_0 = 45\,\mu$m (frame window indicated by a red square in (b)), showing the RTI and the Rayleigh-Plateau instability.

The instability was quantified by defining the spike height as half of the difference between the maximum and minimum radial deviations from the initial drop shape, which was then normalised with the average drop radius, $R_d = 1420\,\mu$m. In figure 20(b), it is evident that the instability is formed immediately after bubble creation, as it grows during the bubble's initial expansion. It starts shrinking at $R_0 \leq 40\,\mu$m during its first collapse and grows again upon its rebound.

For $R_0 = 25.7\,\mu$m, the normalised spike height stays below 0.2 %, meaning that the instability does not further develop in the first bubble oscillation cycles. As the initial radius $R_0$ is increased in steps of $5\,\mu$m, the spike height approximately doubles. Thus, the spike height is exponentially related to the bubble size, i.e. spike height/$R_d \sim e^{R_0 \cdot \text{const.}}$. Considering that the instability increases continuously with increasing laser energy, a threshold for the onset of the instability can only be chosen arbitrarily. Here, we choose an *ad hoc* threshold value as the normalised spike height of 1 % of $R_d$, around which the spike height does not shrink during most of the bubble's first collapse.

Similarly to the observed in the experiments, the simulations of figure 20(b) show how the spikes are ejected earlier in time as the maximum radius reached by the bubble is increased. For $R_0 = 35\,\mu$m, the spikes nearly cross the threshold (indicated in the plot with a dashed line) in the second oscillation cycle, while for $R_0 = 40\,\mu$m, the threshold is crossed shortly after the first collapse, and for $R_0 \geq 45\,\mu$m it is already exceeded during the first oscillation cycle. Figure 20(a) compares the instability for a case below ($R_0 = 35\,\mu$m) and above ($R_0 = 45\,\mu$m) the established threshold, showing a strong increase in the spike size as well as the ejection of droplets for a larger bubble. This droplet separation from the spikes, highlighted in the bottom row of figure 20(c), was previously discussed in figure 18(b) as an example of the effect of the Rayleigh-Plateau instability. An upper limit of the spike height is reached at $\approx 1\,R_d$ for $R_0 = 65\,\mu m$, where the outer spikes reach about twice the drop size, while the inner spikes breach the liquid layer that separates the bubble from the outside air. Because of this, the bubble interior is partially filled with atmospheric gas and the cavity ceases to oscillate. At this point, the drop can not be longer defined as such, as shown in the experiments of figure 19(f) where the liquid mass becomes an intricate collection of spikes and a significant portion of it is ejected away as smaller droplets.

## 5. Conclusion

In this manuscript, we presented some of the complex fluid dynamics occurring once a vapour bubble expands within a water droplet. Specifically, we analysed the appearance of acoustic secondary cavitation, and the formation of liquid jets in the proximity of highly curved free surfaces, and finally, we provided detailed experimental and simulated images of the onset and the development of shape instabilities on the surface of the drop.

The first part of the research highlights that acoustic waves emitted from the micro-explosion nucleate complex secondary cavitation clouds. Further, the study corroborates the existing relation between the evolution of the negative pressure profile and the shape of the bubble clusters inside the drop. A cavitation threshold pressure of around $-4.5\,MPa$ was estimated by performing a direct comparison between the experiments and the simulations. The numerical model does not account for the bubble nucleation induced by the rarefaction waves. The implementation of this experimental technique to other liquids, particularly in cases where large samples are not available, might contribute to achieving a deeper understanding of the nucleation of bubbles by sound waves. The present experimental setup may be modified to create a bubble within a superheated droplet to reveal in a well-defined system the coupling of fluid dynamics with thermodynamics, and also study how the liquid temperature affects the later fragmentation dynamics (Bar-Kohany & Levy 2016).

The secondary bubbles cluster and several types of jets, both caused by the generation of laser bubbles at different positions inside the droplet, were classified using a stand-off parameter $\Upsilon$. The use of a single quantity to characterise the system simplifies the direct comparison between cases. The optical lens effect linked to the spherical shape of the drops allowed us to obtain images of the bubble jet's interior with a remarkable level of detail.

The numerical simulations were crucial to explain the complex flow fields generating these jets, as well as to explain the shape acquired by the gas cavities during their second collapse phase, including many interesting features like the annular bubble necking and the detachment of multiple vapour rings.

The effect of the liquid surface curvature on the bubble jetting has been analysed, by comparing the evolution of a bubble inside a droplet and in a semi-infinite pool, using two complementary points of view. First, a qualitative assessment (here called *behavioural similarity*) indicates that the jetting regime differs rather little when the cavity is seeded nearby the free boundary. In this part of the study, we have shown that for the droplet case the non-dimensional distance $D^*$ is the most determining quantity, while the curvature of the liquid does not have a dominant role in the evolution of the jetting cavities. This conclusion is based in the analysis of numerical simulations where only the parameter $R_d$ was modified, and also by comparing the current results with the previously reported for a flat surface.

A second type of analysis, which uses the CW-SSIM index to evaluate the *structural similarity* of the cavities, was applied on the same numerical data to perform this time a quantitative comparison of the jetting near a flat and a curved surface. Here, we found that for bubbles in the vicinity of the liquid surface (i.e. $0 \lesssim D^* \lesssim 0.7$) the structural similarity is rather poor, mostly due to the higher degree of fragmentation of the gas phase developed in regimes with a ventilated cavity, or where the liquid surface is affected by the RTI.

Both similarity criteria indicate the existence of a seeding depth around $D^* \sim 1$ where the bubbles in the flat and curved cases resemble each other the most. In addition, as the bubble seeding position is set further away from the surface, the jetting regimes are progressively dissimilar, in particular when the laser cavity is generated nearby the drop centre. The sudden drop in the CW-SSIM index found in this situation matches an equally abrupt rise in the value of $\Upsilon$ starting around $\Upsilon = 10$.

The CW-SSIM analysis confirmed that for each stand-off distance in the flat boundary case (i.e. $D_{\text{flat}}^*$), there is another value of $D^*$ (i.e. $D_{\text{drop}}^*$) where the bubble dynamics of both cases resemble each other the most. The relation between $D_{\text{flat}}^*$ and $D_{\text{drop}}^*$ supports the definition and the functionality described for $\Upsilon$. This kind of similarity study could be used to span a more comprehensive parameter space with $D_{\text{flat}}^*$ and $\Upsilon$ computed with different curvature radii, and thus achieve a more general picture of the group of parameter values having "equivalent" jetting dynamics. Moreover, the jet matching would greatly benefit from the implementation of more complex comparison methods or the use of machine learning techniques that consider both the behavioral and the structural criteria.

The spherical bubble oscillations observed in the experiments where the laser was focused on the geometrical centre of the droplet were analysed using two different numerical models. Both models were in excellent agreement with the measured temporal bubble radius evolution. More importantly, both models predict a reduction in the expansion/collapse time when the drop size is decreased. Of course, this study is valid as long as the liquid layer around the bubble is not thin enough to promote the onset of the RTI, as it happens in cases with a low $R_d/R_{max}^*$ ratio.

The radial oscillations of a central bubble were also used to study the onset of shape instabilities at the gas-liquid interfaces, given by the Rayleigh-Taylor and Rayleigh-Plateau instabilities. The destabilisation mechanism of each instability and its effect on the droplet surface was illustrated by detailed high-speed images. Here, we have demonstrated how the radial acceleration imposed by the bubble oscillation triggers the RTI, which in turn induces a pattern of superficial ripples on the drop. Those acquire a concave shape during the bubble collapse and give rise to liquid filaments due to the transfer of the momentum from the bubble shock wave emissions to the curved pits formed on the gas-liquid interface. The ejected filaments later break up by the action of the RPI causing the detachment of smaller droplets and thus the atomisation of the drop.

The phase change from liquid to vapour within droplets is observed in a wide variety of applications, such as in flash boiling atomisation (Loureiro *et al.* 2021), in spray-flame synthesis (Jüngst *et al.* 2022), spray cooling (Tran *et al.* 2012), extreme ultraviolet light generation (Versolato 2019), and laser-induced breakdown spectroscopy of liquids (Lazic & Jovićević 2014) to name a few. They all have in common that through a complex non-spherical symmetric process a liquid is fragmented through a micro-explosion within. While Rayleigh-Taylor instabilities determine the growth of ripples on the surface of the droplet, the non-spherical bubble dynamics that leads to jetting out of the droplet affects the resulting size distribution of liquid particles, too. The high degree of control achieved in the current experiments opens up the possibility of studying RTI of more complex interfaces, e.g. the effect of particles covering the surface, surfactants, or complex fluids. Those experiments could be supported by complementary numerical simulations to optimise the work flow in the laboratory.

## Funding

J.M.R and K.A.R. acknowledge support by the Alexander von Humboldt Foundation (Germany) through the Georg Forster and Humboldt Research Fellowships. This project has received funding from the European Union's Horizon research and innovation programme under the Marie Sk lodowska-Curie grant agreement No. 101064097, as well as the Deutsche Forschungsgemeinschaft (DFG, German Research Foundation) under contract OH75/4-1 and INST 272/280-1. The authors would also like to thank the anonymous reviewers for their constructive criticism and suggestions that helped to improve this work.

## Declaration of interests

The authors report no conflict of interest.

## Author contributions

**Juan Manuel Rosselló**: Conceptualisation, Data curation, Formal Analysis, Funding acquisition, Investigation, Methodology, Software, Visualisation and Writing – original draft, Writing – review & editing. **Hendrik Reese**: Conceptualisation, Data curation, Formal Analysis, Software, Visualisation and Writing – review & editing. **K. Ashoke Raman**: Conceptualisation, Funding acquisition and Investigation. **Claus-Dieter Ohl**: Conceptualisation, Formal Analysis, Funding acquisition, Resources and Writing – review & editing.

## Appendix A. Bubble jetting in a liquid pool with a curved free surface

In section 4.2, the role of the curvature of the free surface was analysed by comparing the bubble jetting observed near a flat surface to the jetting dynamics of bubbles within the falling droplet. There, both the experimental and numerical results indicate that the effect of the curvature on the bubble jetting regime is almost negligible for low values of $D^*$, but the specific shape acquired by the cavity during and after the jetting are no longer similar as $D^*$ takes values larger than 1.2, as shown by the structural similarity analysis. However, the curvature of the surface is not the only difference between these two cases, since in one case the liquid is confined (i.e. the droplet) and in the other the bubble is produced on top of an ideally "semi-infinite" liquid column (which has a length of 5 cm in the experiments and was numerically infinite in the simulations). An intermediate step between those two experimental scenarios is given by the configuration described in figure 21(a). Here, the bubbles are also produced close to the free surface of a liquid pool, but in this case the top of the liquid column presents a curved surface with the shape of a dome. Panels (b), (c) and (d) of figure 21 show three examples of jetting bubbles generated at different depths $d$. The bubbles were located away from the symmetry axis to make evident that the jets always point in the direction normal to the surface. As discussed in section 4.2, we observed a similar behaviour of the jetting dynamics for both curved surfaces at comparable values of the stand-off parameter. The example shown in panel (b) of figure 21 corresponds to the case (c) of figure 12, while the jet dynamics of the case (c) of figure 21 matches the one of figure 12(a).

Figure 22 compares the jetting of bubbles in the drop case with the jetting of bubbles in a configuration as the one shown in figure 21(a). The results present almost identical bubble dynamics even in the case with $D^* = 1.23$, meaning that the differences observed between the case with the flat surface and the drop case are indeed caused by the effect of the surface curvature and not by the difference in the boundary conditions below the bubble or in the liquid volume. As previously discussed for the flat surface case, the similarity found in the cases displayed in figure 22 will be eventually lost as the laser bubble is produced closer to the drop centre.

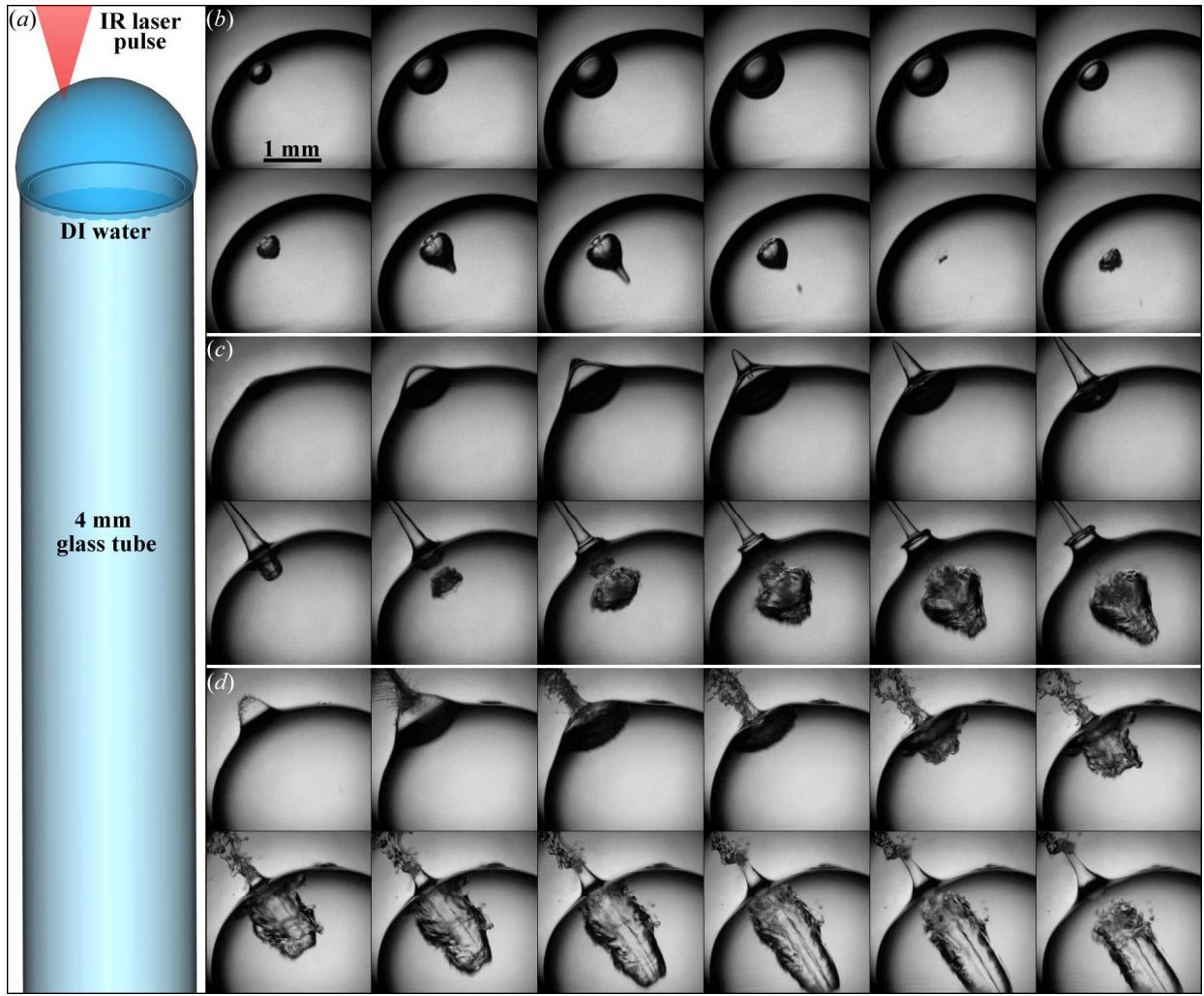

Figure 21: Bubble jetting on a curved surface. (a) A cylindrical tube with an external diameter of 3.6 mm, an internal diameter of 2.7 mm and a length of 4 cm was overfilled with DI water to produce the curved top. Infra-red laser pulses were focused from the top at different depths and slightly away from the cylinder axis. (b) $¥ = 1.9$ ($D^* = 1.3$). (c) $¥ = 0.7$ ($D^* = 0.6$). (d) $¥ = 0.4$ ($D^* = 0.3$). The time between frames is 10 µs for (b) and (c) and 20 µs in case (d).

## REFERENCES

Aganin, A. A., Kosolapova, L. A. & Malakhov, V. G. 2022 Bubble dynamics near a locally curved region of a plane rigid wall. *Physics of Fluids* **34** (9), 097105.

Alexander, D. R. & Armstrong, J. G. 1987 Explosive vaporization of aerosol drops under irradiation by a CO$_2$ laser beam. *Applied Optics* **26** (3), 533.

Ando, K., Liu, A. & Ohl, C. D. 2012 Homogeneous Nucleation in Water in Microfluidic Channels. *Physical Review Letters* **109** (4), 044501.

Andrews, E. D., Fernández Rivas, D. & Peters, I. R. 2020 Cavity collapse near slot geometries. *Journal of Fluid Mechanics* **901** (June).

Andrews, E. D. & Peters, I. R. 2022 Modeling bubble collapse anisotropy in complex geometries. *Phys. Rev. Fluids* **7**, 123601.

Atchley, A.A., Frizzell, L.A., Apfel, R.E., Holland, C.K., Madanshetty, S. & Roy, R.A. 1988 Thresholds for cavitation produced in water by pulsed ultrasound. *Ultrasonics* **26** (5), 280–285.

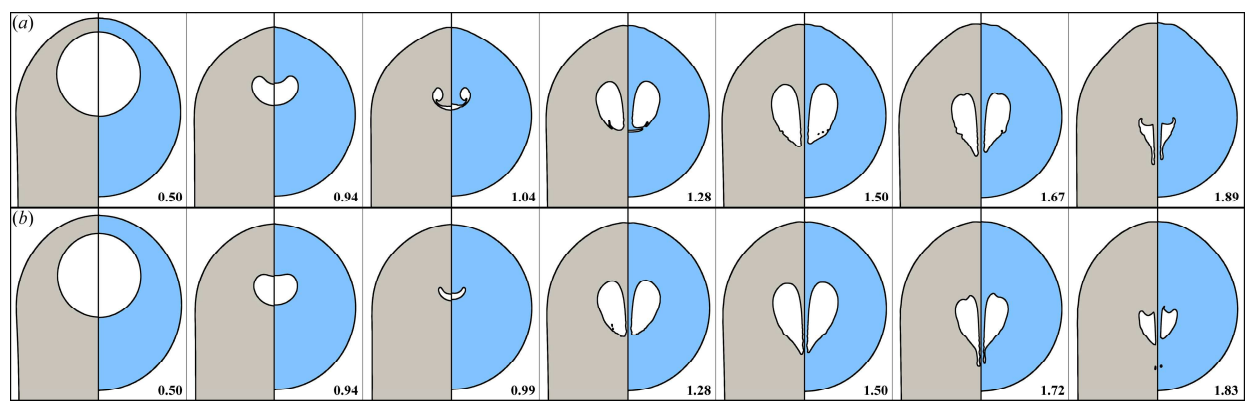

Figure 22: Comparison of the bubble jetting on the curved top of a long liquid column (left half) and in a drop of equal radius (right half). The numbers represent the time normalised with the time of collapse of the cavities from each case. (a) $D^* = 1.02$. (b) $D^* = 1.23$. The dynamics of the bubbles are almost identical.

Bao, H., Reuter, F., Zhang, H., Lu, J. & Ohl, C. D. 2023 Impact-Driven Cavitation Bubble Dynamics. *Experiments in Fluids* **64**, 27.

Bar-Kohany, T. & Levy, M. 2016 State of the art review of flash-boiling atomization. *Atomization and Sprays* **26** (12).

Bempedelis, N., Zhou, J., Andersson, M. & Ventikos, Y. 2021 Numerical and experimental investigation into the dynamics of a bubble-free-surface system. *Physical Review Fluids* **6** (1), 013606.

Biasiori-Poulanges, L. & Schmidmayer, K. 2023 A phenomenological analysis of droplet shock-induced cavitation using a multiphase modeling approach. *Physics of Fluids* **35** (1).

Blake, J. R., Keen, G. S., Tong, R. P. & Wilson, M. 1999 Acoustic cavitation: the fluid dynamics of non–spherical bubbles. *Philosophical Transactions of the Royal Society of London. Series A: Mathematical, Physical and Engineering Sciences* **357** (1751), 251–267.

Blake, J. R., Leppinen, D. M. & Wang, Q. 2015 Cavitation and bubble dynamics: the Kelvin impulse and its applications. *Interface Focus* **5** (5), 20150017.

Brujan, E. A., Keen, G. S., Vogel, A. & Blake, J. R. 2002 The final stage of the collapse of a cavitation bubble close to a rigid boundary. *Physics of Fluids* **14** (1), 85–92.

Brujan, E. A., Nahen, K., Schmidt, P. & Vogel, A. 2001 Dynamics of laser-induced cavitation bubbles near elastic boundaries: influence of the elastic modulus. *Journal of Fluid Mechanics* **433**, 283–314.

Brujan, E. A., Zhang, A., Liu, Y., Ogasawara, T. & Takahira, H. 2022 Jetting and migration of a laser-induced cavitation bubble in a rectangular channel. *Journal of Fluid Mechanics* **948** (September), A6.

Eickmans, J. H., Hsieh, W. F. & Chang, R. K. 1987 Laser-induced explosion of $H_2O$ droplets: spatially resolved spectra. *Optics Letters* **12** (1), 22.

Favre, C., Boutou, V., Hill, S. C., Zimmer, W., Krenz, M., Lambrecht, H., Yu, J., Chang, R. K., Woeste, L. & Wolf, J. 2002 White-Light Nanosource with Directional Emission. *Physical Review Letters* **89** (3), 035002.

Gonzalez-Avila, S. R., Denner, F. & Ohl, C. D. 2021 The acoustic pressure generated by the cavitation bubble expansion and collapse near a rigid wall. *Physics of Fluids* **33** (3), 032118.

Gonzalez-Avila, S. R., Klaseboer, E., Khoo, B. C. & Ohl, C. D. 2011 Cavitation bubble dynamics in a liquid gap of variable height. *Journal of Fluid Mechanics* **682** (September), 241–260.

Gonzalez-Avila, S. R. & Ohl, C. D. 2016 Fragmentation of acoustically levitating droplets by laser-induced cavitation bubbles. *Journal of Fluid Mechanics* **805**, 551–576.

Gonzalez-Avila, S. R., Song, C. & Ohl, C. D. 2015 Fast transient microjets induced by hemispherical cavitation bubbles. *Journal of Fluid Mechanics* **767**, 31–51.

Grünbein, M. L., Gorel, A., Foucar, L., Carbajo, S., Colocho, W., Gilevich, S., Hartmann, E., Hilpert, M., Hunter, M., Kloos, M., Koglin, J. E., Lane, T. J., Lewandowski, J., Lutman, A., Nass, K., Nass Kovacs, G., Roome, C. M., Sheppard, J., Shoeman, R. L., Stricker, M., van Driel, T., Vetter, S., Doak, R. B., Boutet, S., Aquila, A., Decker, F. J., Barends, T. R. M., Stan, C. A. & Schlichting, I. 2021 Effect of X-ray free-electron laser-induced shockwaves on haemoglobin microcrystals delivered in a liquid jet. *Nature Communications* **12** (1), 1672.

HAGEMANN, J., VASSHOLZ, M., HOEPPE, H., OSTERHOFF, M., ROSSELLÓ, J. M., METTIN, R., SEIBOTH, F., SCHROPP, A., MÖLLER, J., HALLMANN, J., KIM, C., SCHOLZ, M., BOESENBERG, U., SCHAFFER, R., ZOZULYA, A., LU, W., SHAYDUK, R., MADSEN, A., SCHROER, C. G. & SALDITT, T. 2021 Single-pulse phase-contrast imaging at free-electron lasers in the hard X-ray regime. *Journal of Synchrotron Radiation* **28** (1), 52–63.

HEIJNEN, L., QUINTO-SU, P. A., ZHAO, X. & OHL, C. D. 2009 Cavitation within a droplet. *Physics of Fluids* **21** (9).

JÜNGST, N., SMALLWOOD, G. J. & KAISER, S. A. 2022 Visualization and image analysis of droplet puffing and micro-explosion in spray-flame synthesis of iron oxide nanoparticles. *Experiments in Fluids* **63**, 60.

KADIVAR, E., PHAN, T., PARK, W. & EL MOCTAR, O. 2021 Dynamics of a single cavitation bubble near a cylindrical rod. *Physics of Fluids* **33** (11), 113315.

KELLER, J. B. & KOLODNER, I. 1954 Instability of Liquid Surfaces and the Formation of Drops. *Journal of Applied Physics* **25** (7), 918–921.

KIYAMA, A., SHIMAZAKI, T., GORDILLO, J. M. & TAGAWA, Y. 2021 Direction of the microjet produced by the collapse of a cavitation bubble located in a corner of a wall and a free surface. *Phys. Rev. Fluids* **6**, 083601.

KLEIN, A. L., KURILOVICH, D., LHUISSIER, H., VERSOLATO, O. O., LOHSE, D., VILLERMAUX, E. & GELDERBLOM, H. 2020 Drop fragmentation by laser-pulse impact. *Journal of Fluid Mechanics* **893**, A7.

KOCH, M., ROSSELLÓ, J. M., LECHNER, C., LAUTERBORN, W., EISENER, J. & METTIN, R. 2021*a* Theory-assisted optical ray tracing to extract cavitation-bubble shapes from experiment. *Experiments in Fluids* **62** (3), 60.

KOCH, M., ROSSELLÓ, J. M., LECHNER, C., LAUTERBORN, W. & METTIN, R. 2021*b* Dynamics of a laser-induced bubble above the flat top of a solid cylinder—mushroom-shaped bubbles and the fast jet. *Fluids* **7** (1), 2.

KONDO, T. & ANDO, K. 2016 One-way-coupling simulation of cavitation accompanied by high-speed droplet impact. *Physics of Fluids* **28** (3), 033303.

KOUKOUVINIS, P., GAVAISES, M., SUPPONEN, O. & FARHAT, M. 2016 Simulation of bubble expansion and collapse in the vicinity of a free surface. *Physics of Fluids* **28** (5), 052103.

KYRIAZIS, N., KOUKOUVINIS, P. & GAVAISES, M. 2018 Modelling cavitation during drop impact on solid surfaces. *Advances in Colloid and Interface Science* **260**, 46–64.

LAUTERBORN, W. & BOLLE, H. 1975 Experimental investigations of cavitation-bubble collapse in the neighbourhood of a solid boundary. *Journal of Fluid Mechanics* **72** (02), 391.

LAUTERBORN, WERNER, LECHNER, CHRISTIANE, KOCH, MAX & METTIN, ROBERT 2018 Bubble models and real bubbles: Rayleigh and energy-deposit cases in a Tait-compressible liquid. *IMA Journal of Applied Mathematics* **83** (4), 556–589.

LAZIC, V. & JOVIĆEVIĆ, S. 2014 Laser induced breakdown spectroscopy inside liquids: Processes and analytical aspects. *Spectrochimica Acta Part B: Atomic Spectroscopy* **101**, 288–311.

LECHNER, C., KOCH, M., LAUTERBORN, W. & METTIN, R. 2017 Pressure and tension waves from bubble collapse near a solid boundary: A numerical approach. *The Journal of the Acoustical Society of America* **142** (6), 3649–3659.

LEE, H., PARTANEN, M., LEE, M., JEONG, S., LEE, H. JOO, KIM, K., RYU, W., DHOLAKIA, K. & OH, K. 2022 A laser-driven optical atomizer: photothermal generation and transport of zeptoliter-droplets along a carbon nanotube deposited hollow optical fiber. *Nanoscale* **14**, 5138–5146.

LI, S., ZHANG, A., HAN, R. & MA, Q. 2019*a* 3D full coupling model for strong interaction between a pulsating bubble and a movable sphere. *Journal of Computational Physics* **392**, 713–731.

LI, S., ZHANG, A. M., HAN, R. & LIU, Y. Q. 2017 Experimental and numerical study on bubble-sphere interaction near a rigid wall. *Physics of Fluids* **29** (9).

LI, S. M., ZHANG, A. M., WANG, Q. X. & ZHANG, S. 2019*b* The jet characteristics of bubbles near mixed boundaries. *Physics of Fluids* **31** (10), 107105.

LI, T., ZHANG, A. M., WANG, S. P., LI, S. & LIU, W. T. 2019*c* Bubble interactions and bursting behaviors near a free surface. *Physics of Fluids* **31** (4), 042104.

LIANG, X. X., LINZ, N., FREIDANK, S., PALTAUF, G. & VOGEL, A. 2022 Comprehensive analysis of spherical bubble oscillations and shock wave emission in laser-induced cavitation. *Journal of Fluid Mechanics* **940**, 1–56.

Lindau, O. & Lauterborn, W. 2003 Cinematographic observation of the collapse and rebound of a laser-produced cavitation bubble near a wall. *Journal of Fluid Mechanics* **479**, 327–348.

Lindinger, A., Hagen, J., Socaciu, L. D., Bernhardt, T. M., Wöste, L., Duft, D. & Leisner, T. 2004 Time-resolved explosion dynamics of $H_2O$ droplets induced by femtosecond laser pulses. *Applied Optics* **43** (27), 5263.

Liu, N. N., Wu, W. B., Zhang, A. M. & Liu, Y. L. 2017 Experimental and numerical investigation on bubble dynamics near a free surface and a circular opening of plate. *Physics of Fluids* **29** (10), 107102.

Loureiro, D.D., Kronenburg, A., Reutzsch, J., Weigand, B. & Vogiatzaki, K. 2021 Droplet size distributions in cryogenic flash atomization. *International Journal of Multiphase Flow* **142**, 103705.

Maeda, K. & Colonius, T. 2019 Bubble cloud dynamics in an ultrasound field. *Journal of Fluid Mechanics* **862**, 1105–1134, arXiv: 1805.00129.

Mahmud, M., Smith, W. R. & Walmsley, A. D. 2020 Numerical investigation of bubble dynamics at a corner. *Physics of Fluids* **32** (5), 053306.

Marston, J. O. & Thoroddsen, S. T. 2015 Laser-induced micro-jetting from armored droplets. *Experiments in Fluids* **56** (7), 140.

Martins, F. J. W. A., da Silva, C. C., Lessig, C. & Zähringer, K. 2018 Ray-tracing based image correction of optical distortion for piv measurements in packed beds. *Journal of Advanced Optics and Photonics* **1** (2), 71–94.

Mei, L. & Brydegaard, M. 2015 Atmospheric aerosol monitoring by an elastic scheimpflug lidar system. *Opt. Express* **23** (24), A1613–A1628.

Mur, J., Agrež, V., Zevnik, J., Petkovšek, R. & Dular, M. 2023 Microbubble collapse near a fiber: Broken symmetry conditions and a planar jet formation. *Physics of Fluids* **35** (2), 023305.

Noack, J. & Vogel, A. 1998 Single-shot spatially resolved characterization of laser-induced shock waves in water. *Appl. Opt.* **37** (19), 4092–4099.

Obreschkow, D., Kobel, P., Dorsaz, N., de Bosset, A., Nicollier, C. & Farhat, M. 2006 Cavitation Bubble Dynamics inside Liquid Drops in Microgravity. *Physical Review Letters* **97** (9), 094502.

OpenFOAM-v2006 2020 https://www.openfoam.com/download/release-history.

Peters, I. R., Tagawa, Y., Oudalov, N., Sun, C., Prosperetti, A., Lohse, D. & van der Meer, D. 2013 Highly focused supersonic microjets: numerical simulations. *Journal of Fluid Mechanics* **719**, 587–605.

Plesset, M. S. & Chapman, R. B. 1971 Collapse of an initially spherical Vapor Cavity in the Neighborhood of a solid Boundary. *Journal of Fluid Mechanics* **47** (2), 283–290.

Quinto-Su, P. A. & Ando, K. 2013 Nucleating bubble clouds with a pair of laser-induced shocks and bubbles. *Journal of Fluid Mechanics* **733**, R3.

Raman, C. V. & Sutherland, G. A. 1922 On the whispering-gallery phenomenon. *Proceedings of the Royal Society of London. Series A, Containing Papers of a Mathematical and Physical Character* **100**, 424–428.

Raman, K. A., Rosselló, J. M., Reese, H. & Ohl, C. D. 2022 Microemulsification from single laser-induced cavitation bubbles. *Journal of Fluid Mechanics* **953**, A27.

Reese, H., Schädel, R., Reuter, F. & Ohl, C. D. 2022 Microscopic pumping of viscous liquids with single cavitation bubbles. *Journal of Fluid Mechanics* **944** (May), A17.

Ren, Z., Zuo, Z., Wu, S. & Liu, S. 2022 Particulate projectiles driven by cavitation bubbles. *Phys. Rev. Lett.* **128**, 044501.

Robert, E., Lettry, J., Farhat, M., Monkewitz, P. A. & Avellan, F. 2007 Cavitation bubble behavior inside a liquid jet. *Physics of Fluids* **19** (6), 067106.

Rohwetter, P., Kasparian, J., Stelmaszczyk, K., Hao, Z., Henin, S., Lascoux, N., Nakaema, W. M., Petit, Y., Queißer, M., Salamé, R., Salmon, E., Wöste, L. & Wolf, J. 2010 Laser-induced water condensation in air. *Nature Photonics* **4** (7), 451–456.

Rosselló, J. M. & Ohl, C. D. 2022 Bullet jet as a tool for soft matter piercing and needle-free liquid injection. *Biomedical Optics Express* **13** (10), 5202.

Rosselló, J. M., Reese, H. & Ohl, C. D. 2022 Dynamics of pulsed laser-induced cavities on a liquid–gas interface: from a conical splash to a 'bullet' jet. *Journal of Fluid Mechanics* **939**, A35.

Sampat, M. P., Wang, Z., Gupta, S., Bovik, A. C. & Markey, M. K. 2009 Complex wavelet structural similarity: A new image similarity index. *IEEE Transactions on Image Processing* **18** (11), 2385–2401.

Sembian, S., Liverts, M., Tillmark, N. & Apazidis, N. 2016 Plane shock wave interaction with a cylindrical water column. *Physics of Fluids* **28** (5), 056102.

SINGH, P. I. & KNIGHT, C. J. 1980 Pulsed Laser-Induced Shattering of Water Drops. *AIAA Journal* **18** (1), 96–100.

STAN, C. A., MILATHIANAKI, D., LAKSMONO, H., SIERRA, R. G., MCQUEEN, T. A., MESSERSCHMIDT, M., WILLIAMS, G. J., KOGLIN, J. E., LANE, T. J., HAYES, M. J., GUILLET, S. A. H., LIANG, M., AQUILA, A. L., WILLMOTT, P. R., ROBINSON, J. S., GUMERLOCK, K. L., BOTHA, S., NASS, K., SCHLICHTING, I., SHOEMAN, R. L., STONE, H. A. & BOUTET, S. 2016 Liquid explosions induced by X-ray laser pulses. *Nature Physics* **12** (10), 966–971.

SUPPONEN, O., OBRESCHKOW, D., TINGUELY, M., KOBEL, P., DORSAZ, N. & FARHAT, M. 2016 Scaling laws for jets of single cavitation bubbles. *Journal of Fluid Mechanics* **802** (2016), 263–293.

TAGAWA, Y., OUDALOV, N., VISSER, C. W., PETERS, I. R., VAN DER MEER, D., SUN, C., PROSPERETTI, A. & LOHSE, D. 2012 Highly focused supersonic microjets. *Phys. Rev. X* **2**, 031002.

TAYLOR, G. I. 1950 The formation of a blast wave by a very intense explosion I. Theoretical discussion. *Proceedings of the Royal Society of London. Series A. Mathematical and Physical Sciences* **201** (1065), 159–174.

THORODDSEN, S. T., TAKEHARA, K., ETOH, T. G. & OHL, C. D. 2009 Spray and microjets produced by focusing a laser pulse into a hemispherical drop. *Physics of Fluids* **21** (11), 112101.

TOMITA, Y., ROBINSON, P. B., TONG, R. P. & BLAKE, J. R. 2002 Growth and collapse of cavitation bubbles near a curved rigid boundary. *Journal of Fluid Mechanics* **466**, 259–283.

TRAN, T., STAAT, H. J. J., PROSPERETTI, A., SUN, C. & LOHSE, D. 2012 Drop impact on superheated surfaces. *Physical Review Letters* **108**, 036101.

TRUMMLER, T., BRYNGELSON, S. H., SCHMIDMAYER, K., SCHMIDT, S. J., COLONIUS, T. & ADAMS, N. A. 2020 Near-surface dynamics of a gas bubble collapsing above a crevice. *Journal of Fluid Mechanics* .

URSESCU, D., ALEKSANDROV, V., MATEI, D., DANCUS, I., DE ALMEIDA, M. D. & STAN, C. A. 2020 Generation of shock trains in free liquid jets with a nanosecond green laser. *Physical Review Fluids* **5** (12), 123402.

VASSHOLZ, M., HOEPPE, H. P., HAGEMANN, J., ROSSELLÓ, J. M., OSTERHOFF, M., METTIN, R., KURZ, T., SCHROPP, A., SEIBOTH, F., SCHROER, C. G., SCHOLZ, M., MÖLLER, J., HALLMANN, J., BOESENBERG, U., KIM, C., ZOZULYA, A., LU, W., SHAYDUK, R., SCHAFFER, R., MADSEN, A. & SALDITT, T. 2021 Pump-probe X-ray holographic imaging of laser-induced cavitation bubbles with femtosecond FEL pulses. *Nature Communications* **12** (1), 3468.

VASSHOLZ, M., HOEPPE, H. P., HAGEMANN, J., ROSSELLÓ, J. M., OSTERHOFF, M., METTIN, R., MÖLLER, J., SCHOLZ, M., BOESENBERG, U., HALLMANN, J., KIM, C., ZOZULYA, A., LU, W., SHAYDUK, R., MADSEN, A. & SALDITT, T. 2023 Structural dynamics of water in a supersonic shockwave. *Physics of Fluids* **35** (1), 016126.

VERSOLATO, O. O. 2019 Physics of laser-driven tin plasma sources of EUV radiation for nanolithography. *Plasma Sources Science and Technology* **28** (8), 083001.

VOGEL, A., BUSCH, S. & PARLITZ, U. 1996 Shock wave emission and cavitation bubble generation by picosecond and nanosecond optical breakdown in water. *The Journal of the Acoustical Society of America* **100** (1), 148–165.

WANG, S. P., WANG, Q. X., LEPPINEN, D. M., ZHANG, A. M. & LIU, Y. L. 2018 Acoustic bubble dynamics in a microvessel surrounded by elastic material. *Physics of Fluids* **30** (1), 012104.

WU, S., ZUO, Z., REN, Z. & LIU, S. 2018a Dynamics of a Laser-induced Bubble Near a Convex Free Surface. In *Proceedings of the 10th International Symposium on Cavitation (CAV2018)*. ASME Press.

WU, W., LIU, Q. & WANG, B. 2021 Curved surface effect on high-speed droplet impingement. *Journal of Fluid Mechanics* **909** (11402298), A7.

WU, W., XIANG, G. & WANG, B. 2018b On high-speed impingement of cylindrical droplets upon solid wall considering cavitation effects. *Journal of Fluid Mechanics* **857**, 851–877.

YANG, Y. X., WANG, Q. X. & KEAT, T. S. 2013 Dynamic features of a laser-induced cavitation bubble near a solid boundary. *Ultrasonics Sonochemistry* **20** (4), 1098–1103.

ZENG, Q., GONZALEZ-AVILA, S. R., TEN VOORDE, S. & OHL, C. D. 2018 Jetting of viscous droplets from cavitation-induced Rayleigh-Taylor instability. *Journal of Fluid Mechanics* **846**, 916–943.

ZEVNIK, J. & DULAR, M. 2020 Cavitation bubble interaction with a rigid spherical particle on a microscale. *Ultrasonics Sonochemistry* **69** (April), 105252.

ZHANG, A., LI, S., CUI, P., LI, S. & LIU, Y. 2023 A unified theory for bubble dynamics. *Physics of Fluids* **35** (3), 033323.

ZHANG, Y., CHEN, F., ZHANG, Y., ZHANG, Y. & DU, X. 2018 Experimental investigations of interactions

3636

between a laser-induced cavitation bubble and a spherical particle. *Experimental Thermal and Fluid Science* **98**, 645–661.

ZHANG, Y., QIU, X., ZHANG, X., TANG, N. & ZHANG, Y. 2020 Collapsing dynamics of a laser-induced cavitation bubble near the edge of a rigid wall. *Ultrasonics Sonochemistry* **67** (October 2019), 105157.

ZHOU, W. & SIMONCELLI, E.P. 2005 Translation insensitive image similarity in complex wavelet domain. In *Proceedings. (ICASSP '05). IEEE International Conference on Acoustics, Speech, and Signal Processing, 2005.*, , vol. 2, pp. ii/573–ii/576 Vol. 2.

ZHOU, Y. 2017*a* Rayleigh–taylor and richtmyer–meshkov instability induced flow, turbulence, and mixing. i. *Physics Reports* **720-722**, 1–136.

ZHOU, Y. 2017*b* Rayleigh–taylor and richtmyer–meshkov instability induced flow, turbulence, and mixing. ii. *Physics Reports* **723-725**, 1–160.